\documentclass[aps,prd,preprint,preprintnumbers,unsortedaddress,superscriptaddress,showpacs,nofootinbib]{revtex4-1}

\pdfoutput=1

\usepackage{graphicx}
\usepackage{amsmath}
\usepackage{amsfonts}
\usepackage{amssymb}
\usepackage{color}

\usepackage[colorlinks, citecolor=blue,anchorcolor=red,menucolor=red, linkcolor=red,filecolor=red,runcolor=red,urlcolor=blue,frenchlinks=red]{hyperref}

\usepackage{bm}
\usepackage[section]{placeins}
\usepackage{booktabs}

\begin{document}
\arraycolsep1.5pt

\title{Semileptonic decays  of  $B^{(*)}$, $D^{(*)}$  into ${\nu} l$  and pseudoscalar or vector mesons }

\author{L.~R.~Dai}
\email{dailr@lnnu.edu.cn}
\affiliation{Department of Physics, Liaoning Normal University, Dalian 116029, China}
\affiliation{Departamento de F\'isica Te\'orica and IFIC, Centro Mixto Universidad de Valencia-CSIC,
Institutos de Investigac\'ion de Paterna, Aptdo. 22085, 46071 Valencia, Spain
}

\author{X. Zhang}
\affiliation{Departamento de F\'isica Te\'orica and IFIC, Centro Mixto Universidad de Valencia-CSIC,
Institutos de Investigac\'ion de Paterna, Aptdo. 22085, 46071 Valencia, Spain
}
\affiliation{Institute of modern physics, Chinese Academy of
Sciences, Lanzhou 730000, China}
\affiliation{University of Chinese Academy of Sciences, Beijing 101408, China}

\author{E. Oset}
\email{oset@ific.uv.es}
\affiliation{Departamento de F\'isica Te\'orica and IFIC, Centro Mixto Universidad de Valencia-CSIC,
Institutos de Investigac\'ion de Paterna, Aptdo. 22085, 46071 Valencia, Spain
}

\date{\today}
\begin{abstract}
 We perform a study of the $B^{(*)}$, $D^{(*)}$  semileptonic decays, using a different method than in conventional approaches, where the matrix elements
 of the  weak operators  are evaluated and a detailed spin-angular momentum algebra is performed  to obtain very simple  expressions at the end  for the different decay modes.
 Using only one experimental decay rate in the $B$ or $D$ sectors, the rates for the rest of decay modes are predicted  and they are in good agreement with experiment. Some
 discrepancies  are observed in the $\tau$ decay mode for which we  find an explanation. We perform  evaluations for $B^*$ and  $D^*$ decay rates that can be used in future
 measurements, now possible in the LHCb collaboration.
 \end{abstract}

\maketitle

\section{Introduction}
\label{intro}
Semileptonic decays of mesons have been thoroughly studied, and are a source of information on the
Cabibbo-Kobayashi-Maskawa (CKM) matrix elements \cite{browder,isgur,isgur2,Wirbel:1988ft,neubert},
chiral dynamics \cite{ecker}, heavy quark symmetry \cite{neubert2}. The process is relatively well understood,
to the point that some discrepancies seen in ratios of rates are proposed as signals of new physics \cite{Antonelli:2009ws,Fajfer,German}.
Concerning the decays of mesons with heavy flavors, the decay of $\bar B \to D \bar \nu l^-$ and $D \to \bar K \nu l^+$ and the
related reactions with $B^*$ or $D^*$ offer a good ground to study heavy flavor symmetry.

In the conventional approaches the amplitudes of the processes are conveniently parameterized
in terms of certain structures and their associated form factors, and some information is taken
from experiment. Quark models can provide information on these form factors and structures and have been often used \cite{isgur,isgur2,nieves}.

  The purpose of this paper is to see how far one can go, assuming basic facts of heavy quark symmetry, with some caution that will be
  discussed later, which allows us to conclude that the relevant form factors would be the same for $D$ or $D^*$ and $B$ or $B^*$.
  Yet, the structures can be very different due to the angular momentum combinations that the quarks undergo to produce the pseudoscalar
  or vector meson states. This is what is accomplished in the present work, where a detailed study is done of the amplitudes for each of the four
  $B^{(*)} \to D^{(*)} \bar \nu l^-$ cases, and the corresponding ones with $D^{(*)}$, evaluating explicitly the weak matrix elements in the rest frame
  of the $ \nu l$ and performing the angular momentum algebra which relates all the processes. We then fit the results to one experimental
  branching ratio for the $B$ and $D$ sectors and then the rest of the results are predictions.

 The derivation requires some patience, but we succeed using Racah algebra to write the final amplitudes and the sums over polarizations
 of their modulus square in terms of very simple analytical expressions, which allow us to explain easily some of the features of the
 reactions, as the relative rates in the different sectors and peculiarities of the differential width distribution in the invariant
 mass of the  $ \nu l$ system.

  One of the outputs of the work is the prediction of rates for $B^*$ and $D^*$ decays, which have received
  attention recently \cite{wangwang,changyang} in view of the possibility that such decay rates are observed by the LHCb collaboration.
  We argue that the method proposed is highly accurate to make predictions for these decay widths.

   As to the predictions for the observed rates, the method is rather accurate for the case of production of light leptons,
   and has some discrepancy for the production of $\tau$ lepton, for which a justification is given, but even then ratios
   of different decay rates with $\tau$ leptons in the final state are also well reproduced.

\section{Formalism}
\label{sec:form}
\begin{figure}[ht]
\includegraphics[scale=0.8]{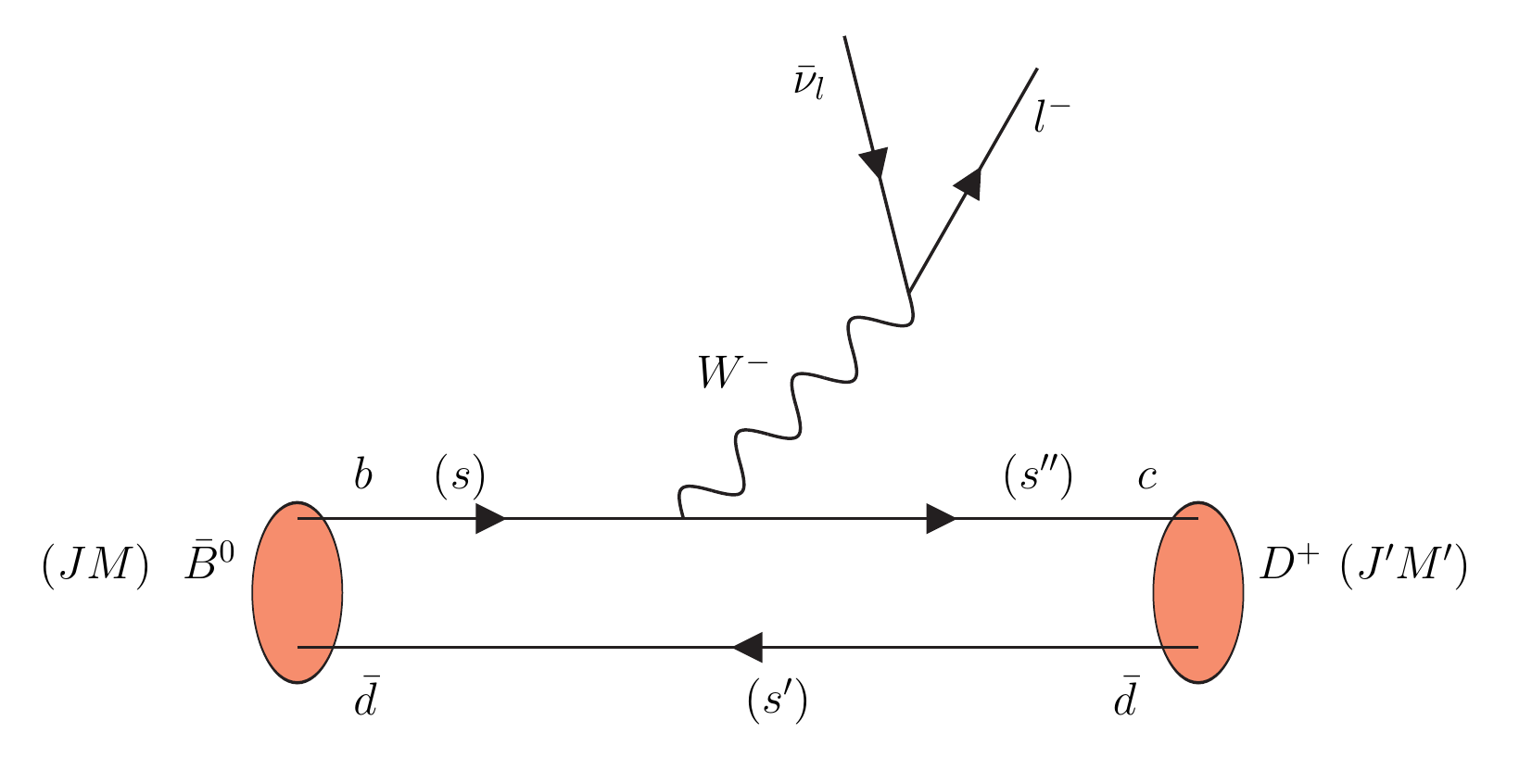}
\caption{Diagrammatic representation of  $\bar{B}^0 \to \bar{\nu}_{l} l^- D^+$  at the quark level.}
\label{fig:diag}
\end{figure}

We shall study reactions of the type $\bar{B}^0 \to \bar{\nu}_{l} l^- D^+$, or  $D^+ \to {\nu}_{l} l^+ \bar{K}^0 $  and the corresponding
ones with vector mesons, with the  aim of relating them assuming that the form factors do not change practically when changing $B \to B^*$  or
$D \to D^*$, which is the essence  of heavy quark symmetry \cite{Neubert,Manohar}.  Work on these reactions assuming this symmetry is done in Ref.\cite{nieves}.
The process is depicted  in Fig. \ref{fig:diag}  for the  $\bar{B}^0 \to \bar{\nu}_{l} l^- D^+$

The weak interaction is given by the Hamiltonian
\begin{eqnarray}
H= \mathcal{C} L^\alpha Q_\alpha \,,
\end{eqnarray}
where in  $\mathcal{C}$ one has the couplings of the weak  interaction, but, since we are only concerned about ratios of rates, it
plays no role in our study. The  leptonic current is given by
\begin{eqnarray}
L^\alpha=\langle {\bar u}_l |\gamma^\alpha(1-\gamma_5)| v_\nu\rangle \,,
\end{eqnarray}
and $Q^\alpha$, the quark current, by
\begin{eqnarray}
Q^\alpha=\langle {\bar u}_c|\gamma^\alpha(1-\gamma_5)|u_b\rangle \,.
\end{eqnarray}
In order to obtain the $\bar{B}^0$ decay width we need
\begin{eqnarray}\sum _{\rm lep~ pol} L^\alpha {L^\beta}^{*} ~
\overline{\sum_{quark}}\sum_{\rm pol}  Q_\alpha Q_\beta^{*}\equiv {L}^{\alpha\beta} \overline\sum\sum Q_\alpha Q_\beta^{*} \,,
\end{eqnarray}
where ${L}^{\alpha\beta}$ stands for ${\sum _{\rm pol}} L^\alpha {L^\beta}^{*}$  and is easily evaluated  with the result \cite{Navarra}
\begin{eqnarray}\label{eq:ab}
{L}^{\alpha\beta} = 2~ \frac{p_\nu^\alpha p_l^\beta+p_l^\alpha p_\nu^\beta-p_\nu \cdot p_l g^{\alpha\beta}-i \epsilon^{\rho\alpha\sigma\beta} p_{\nu\rho} p_{l\sigma}}{m_\nu m_l} \,,
\end{eqnarray}
where we adopt the Mandl and Shaw normalization  for fermions \cite{mandl}.  The mass of the neutrino and the lepton  get cancelled in the final formula of the width.

 In Ref. \cite{Navarra} a similar sum and an average  and sum over the  quark spin third components was done. Here  we pay a special attention  to the vector  or pseudoscalar components,
 and the coupling  of spins to given quantum  numbers has to be done prior to the sums over the  third components in the final  $Q_\alpha Q_\beta^{*}$ term. For this  purpose
  we must evaluate  explicitly the quark current $Q_\alpha$. We use the ordinary spinors \cite{Itzykson}
\begin{equation}\label{eq:wfn}
u_r=
\widetilde{A}\left(
\begin{array}{c}
\chi_r\\ \widetilde{B} {\bm{\sigma}}\cdot {\bm{p}} \, \chi_r
\end{array}
\right)\,; \qquad
\widetilde{A}= \left( \frac{E_p +m}{2\, m} \right)^{1/2}\,;  \qquad \widetilde{B}= \frac{1}{E_p +m}\,,
\end{equation}
where $\chi_r$ are the Pauli bispinors and  $m, p$ and $E_p$ are the  mass, momentum and energy of the quark.
Next we  use \cite{Navarra}
\begin{eqnarray}
\frac{p_b}{m_b}=\frac{p_B}{m_B}\,; \qquad \frac{E_b}{m_b}=\frac{E_B}{m_B}\, ,
\label{eq:ok}
\end{eqnarray}
and the same for the $c$ quark. Theses ratios are related to the velocity  of the quarks or $B$ mesons and neglect the internal motion of the quarks inside the meson. We evaluate
the matrix elements in the frame where  the   ${\bar\nu}  l$  system  is at rest, where  ${\bm{p}}_B={\bm{p}}_D={\bm{p}}$ and both have a sizeable velocity. We have in general
\begin{eqnarray}
p = \frac{\lambda ^{1/2} (m_{in}^2, M_{\rm inv}^{2(\nu l)},  m^2_{fin})}{2 M_{\rm inv}^{(\nu l)}} \, ,
\label{eq:p}
\end{eqnarray}
where $m_{in}$, $m_{fin}$ are the masses of the initial, final mesons in the decay, and $M_{\rm inv}^{(\nu l)}$ is the invariant mass of the $\nu l$ pair.
Using Eq. \eqref{eq:ok} we can  now write
\begin{equation}\label{eq:wfn}
u_r=
A\left(
\begin{array}{c}
\chi_r\\ B {\bm{\sigma}}\cdot {\bm{p}}_B\, \chi_r
\end{array}
\right)\,; \qquad
A= \left( \frac{ \frac{E_B}{m_B} + 1}{2} \right)^{1/2}\,;  \qquad B= \frac{1}{m_B(1+\frac{E_B}{m_B})}\,.
\end{equation}
We also use the  $\gamma^\mu$ representation of Ref. \cite{Itzykson}
\begin{equation}\label{eq:ga}
\gamma^0=
\left(
\begin{array}{cc}
~I~ &~0~ \\
0 & -I
\end{array}
\right); \qquad
\gamma_5=
\left(
\begin{array}{cc}
~0~ &~ I~ \\~I~  &~ 0~
\end{array}
\right); \qquad
\gamma^i=
\left(
\begin{array}{cc}
~0~&\sigma^i\\-\sigma^i &~0~
\end{array}
\right) \, ,~~ i=1,2,3 \, .
\end{equation}
As a consequence we have
\begin{eqnarray}
\gamma^0 -\gamma^0 \gamma_5=\left(
\begin{array}{cc}
~I~ &~-I~ \\
I & -I
\end{array}
\right); \qquad   \gamma^i-\gamma^i \gamma_5=\left(
\begin{array}{cc}
~-\sigma^i~ &~\sigma^i~ \\
-\sigma^i & \sigma^i
\end{array}
\right) \, .
\end{eqnarray}
where we denote the Pauli matrices as $\sigma^i$, $\sigma_i$ (i=1,2,3), but they are the same thing.
\begin{eqnarray}
\langle {\bar u}_c|\gamma^0-\gamma^0\gamma_5|u_b\rangle = A A^{\prime}  \left\{(1+B B^{\prime} p^2) \langle \chi_c | \chi_b \rangle -
(B+B^{\prime})\langle \chi_c | {\bm{\sigma}}\cdot {\bm{p}} |\chi_b \rangle
\right\} \,,
\end{eqnarray}
where $p^2$ stands for ${\bm{p}}^2$ from here on, and $A^{\prime},B^{\prime}$ stand for the $D$ meson.

Similarly,
\begin{eqnarray}
\langle {\bar u}_c|\gamma^i-\gamma^i\gamma_5|u_b\rangle &=& A A^{\prime}  \left\{ -\langle \chi_c | {{\sigma}}^i |\chi_b \rangle +
 B \langle \chi_c | {{\sigma}}^i \, {\bm{\sigma}}\cdot {\bm{p}} |\chi_b \rangle \right. \\ \nonumber \,
&+&  \left. B^{\prime} \langle \chi_c |  {\bm{\sigma}}\cdot {\bm{p}}\, {{\sigma}}^i |\chi_b  \rangle - B B^{\prime}  \langle \chi_c |  {\bm{\sigma}}\cdot {\bm{p}} \,{{\sigma}}^i \, {\bm{\sigma}}\cdot {\bm{p}} |\chi_b\rangle
\right\} \,.
\end{eqnarray}
The explicit calculation is simplified if we take ${\bm{p}}$  in the $z$ direction. Then, considering the spins, $s$,$s^{\prime\prime}$, $s^{\prime}$ (of the $\bar{d}$ in the ${\bm{p}}$ direction)
of Fig. \ref{fig:diag} one has
\begin{eqnarray}
\langle \chi_c | {\bm{\sigma}}\cdot {\bm{p}} |\chi_b \rangle = p \, \langle s^{\prime\prime}| \sigma_z |s \rangle = p\, (-1)^{\frac{1}{2}-s}\delta_{ss^{\prime\prime}}
\equiv p \,\sqrt{3}\, {\cal C}(\frac{1}{2} 1 \frac{1}{2}; s,0,s^{\prime\prime})\delta_{ss^{\prime\prime}} \,,
\end{eqnarray}
where  in the last step we evaluate  the $\sigma_z$ matrix element  using the Wigner-Eckart Theorem.  Hence,
\begin{eqnarray}\label{eq:g0}
\langle {\bar u}_c|\gamma^0-\gamma^0\gamma_5|u_b\rangle \equiv M_0 &=&A A^{\prime}  \left\{ (1+B B^{\prime} p^2)  \delta_{s s^{\prime\prime}}
 -(B+B^{\prime})\,p \,\delta_{s s^{\prime\prime}}\,\sqrt{3}\, {\cal C}(\frac{1}{2} 1 \frac{1}{2}; s,0,s^{\prime\prime}) \right\}   ~~~~\\ \nonumber \,
&=& A A^{\prime}  \delta_{s s^{\prime\prime}}  \left\{ 1+B B^{\prime}  p^2 - (B+B^{\prime})\,p \,\sqrt{3}\, {\cal C}(\frac{1}{2} 1 \frac{1}{2}; s,0,s)
\right\} \,.
\end{eqnarray}
\begin{eqnarray}\label{eq:gi}
\langle {\bar u}_c|\gamma^i-\gamma^i\gamma_5|u_b\rangle \equiv N_i &=&A A^{\prime}  \left\{-\langle s^{\prime\prime}| {{\sigma}}^i |s \rangle
+B\,p \, \langle s^{\prime\prime}| {{\sigma}}^i  \sigma_z  |s \rangle
+B^{\prime}\,p \, \langle s^{\prime\prime}| \sigma_z {{\sigma}}^i  |s \rangle
-B B^{\prime}\,p^2 \, \langle s^{\prime\prime}|\sigma_z {{\sigma}}^i \sigma_z |s \rangle   \right\} ~~~~~\\ \nonumber \,
&=& A A^{\prime} \left\{ -\langle s^{\prime\prime}| {{\sigma}}^i |s \rangle
+B\,p \, \langle s^{\prime\prime}| {{\sigma}}^i |s \rangle (-1)^{\frac{1}{2}-s} + B^{\prime}\,p \, \langle s^{\prime\prime}|  {{\sigma}}^i  |s \rangle (-1)^{\frac{1}{2}-s^{\prime\prime}}   \right. \\ \nonumber \,
&& \left.
-B B^{\prime}  p^2 (-1)^{\frac{1}{2}-s}(-1)^{\frac{1}{2}-s^{\prime\prime}} \langle s^{\prime\prime}| {{\sigma}}^i |s \rangle
\right\} \,.
\end{eqnarray}
If  we keep the covariant form $\gamma_i-\gamma_i\gamma_5$ we have
\begin{eqnarray}    \label{eq:gi2}
\langle {\bar u}_c|\gamma_i-\gamma_i\gamma_5|u_b\rangle =A A^{\prime} \langle s^{\prime\prime}| {{\sigma}}^i |s \rangle
\left\{1-B\,p \, (-1)^{\frac{1}{2}-s} -B^{\prime}\,p \, (-1)^{\frac{1}{2}-s^{\prime\prime}}
+ B B^{\prime} p^2 (-1)^{1-s-s^{\prime\prime}}
\right\} ~~~~~\,.
\end{eqnarray}
In order  to work out the angular momentum algebra it is more convenient to evaluate the spherical component of ${{\sigma}}^i \to {{\sigma}}_\mu, \mu=1,0,-1$ in the last equation, and we define,
$M_0$ for the $\gamma^0-\gamma^0\gamma_5$ matrix element  of Eq. \eqref{eq:g0} and  $N_\mu$ for the matrix element of Eq. \eqref{eq:gi} substituting  ${{\sigma}}^i$ by ${{\sigma}}_\mu$.

The explicit evaluation is done in the Appendix  for the $M_0$ and $N_\mu$ matrix elements and we show  here the results.
\begin{itemize}
\item[A)]  $M_0$  matrix element
\begin{itemize}
\item[1)]  $J=0,J^{\prime}=0$
\begin{equation}
M_0=A A^{\prime} (1+B B^{\prime} \,p^2 \,) \delta_{M 0}\, \delta_{M^{\prime} 0}
\end{equation}
\item[2)] $J=0,J^{\prime}=1$
 \begin{equation}
M_0=-A A^{\prime} (B+B^{\prime}) p \delta_{M 0} \,\delta_{M^{\prime} 0}
\end{equation}
\item[3)]  $J=1,J^{\prime}=0$
 \begin{equation}
M_0=-A A^{\prime} (B+B^{\prime}) p \delta_{M 0} \,\delta_{M^{\prime} 0}
\end{equation}
\item[4)] $J=1,J^{\prime}=1$
 \begin{equation}
M_0=A A^{\prime} \left\{(1+B B^{\prime} p^2) -  \sqrt{2}(B +B^{\prime})\,p \,{\cal C}(1 1 1; M,0,M) \right\} \delta_{M M^{\prime}}
\end{equation}
\end{itemize}
\item[B)]$N_\mu$  matrix element
\begin{itemize}
\item[1)] $J=0,J^{\prime}=0$
 \begin{equation}
N_\mu=-A A^{\prime} (B+B^{\prime}) p \delta_{M 0} \,\delta_{M^{\prime} 0}\,\delta_{\mu 0}
\end{equation}
\item[2)]  $J=0,J^{\prime}=1$
 \begin{equation}
N_\mu=A A^{\prime}\left\{1+BB^{\prime} p^2 \,(-1)^{-M^{\prime}} + \sqrt{2} (Bp+B^{\prime} p (-1)^{-M^{\prime}} ) {\cal C}(1 1 1; M^{\prime},0,M^{\prime}) \right\}  \delta_{\mu,M^{\prime}} \,\delta_{M 0}
\end{equation}
\item[3)]  $J=1,J^{\prime}=0$
\begin{eqnarray}
N_\mu&=&A A^{\prime}\left\{(1+BB^{\prime}\,p^2 \,(-1)^{M})(-1)^{-M} ~~~~\right.\\ \nonumber \,
&-&\left. \sqrt{2}(-1)^{M} (B p+ B^{\prime} p (-1)^{M})  \,{\cal C}(1 1 1; M,0,M) \right\}  \delta_{\mu M} \,\delta_{M^{\prime} 0}
 \end{eqnarray}
\item[4)]  $J=1,J^{\prime}=1$
\begin{eqnarray}
N_\mu&=&A A^{\prime}\left\{(1+BB^{\prime}p^2(-1)^{M-M'}) \sqrt{2}\,{\cal C}(1 1 1; M,M'-M,M') -(B p+ B^{\prime} p(-1)^{M-M'})\delta_{M0} ~~~~~\right.\\ \nonumber \,
&-&\left.2 (B p+ B^{\prime} p (-1)^{M-M'}) \,{\cal C}(1 1 1; M'-M,M,M') \,{\cal C}(1 1 1; 0,M,M) \right\}  \delta_{\mu, M^{\prime}-M}
 \end{eqnarray}
 \end{itemize}
 \item[C)] Next we must evaluate ${L}^{\alpha\beta} Q_\alpha Q_\beta^{*}$ and sum  and average  over polarizations. In terms of the  $M_0$ and $N_i$  terms defined before we have the
 combination
 \begin{equation} \label{eq:tt}
\overline{\sum} \sum \left|t\right|^2= L^{00} M_0~ M^{*}_0 + L^{0i}   M_0 ~N^{*}_i + L^{i0}   N_i ~M_0^{*} + L^{ij}  N_i ~N_j^{*}  \,.
\end{equation}
The explicit evaluation is done in the Appendix and  we show here the final results.
\begin{itemize}
\item[1)] $J=0,J^{\prime}=0$
\begin{eqnarray}\label{eq:t00}
\overline{\sum} \sum |t|^2 &=& \frac{(A A')^2}{m_{\nu} m_l} \left\{ \frac{m^2_l(M_{\rm inv}^{2(\nu l)}-m^2_l)}{M_{\rm inv}^{2(\nu l)}} \left(1+BB^{\prime}\,p^2\right)^2   \right.\\ \nonumber \,
&+& \left. 2 \,\left(\widetilde{E}_\nu \widetilde{E}_l +\frac{1}{3} \widetilde{p}_\nu^2 \right) (B+B')^2 p^2 \right\}  \, ,
 \end{eqnarray}
where the magnitudes with tilda are evaluated in the $\nu l$ rest frame.  Thus
\begin{eqnarray}
\widetilde{E}_l=\frac{M_{\rm inv}^{2(\nu l)}+ m^2_l-m^2_\nu}{2\,M_{\rm inv}^{(\nu l)}}  \nonumber \,,  \\
\widetilde{E}_\nu=\frac{M_{\rm inv}^{2(\nu l)}+ m^2_\nu -m^2_l}{2\,M_{\rm inv}^{(\nu l)}} \nonumber \,,  \\
\widetilde{p}_\nu=\frac{\lambda^{1/2}(M_{\rm inv}^{2(\nu l)}, m^2_\nu,m^2_l)}{2 M_{\rm inv}^{(\nu l)}} \,.
\end{eqnarray}
\item[2)] $J=0,J^{\prime}=1$
\begin{eqnarray}\label{eq:t01}
\overline{\sum} \sum |t|^2 &=& \frac{(A A')^2}{m_{\nu} m_l} \left\{ \frac{m^2_l(M_{\rm inv}^{2(\nu l)}-m^2_l)}{M_{\rm inv}^{2(\nu l)}} (B+B^{\prime})^2 p^2  \right.\\ \nonumber \,
&+& \left. 2 \,\left(\widetilde{E}_\nu \widetilde{E}_l +\frac{1}{3} \widetilde{p}^2_\nu \right) \left(3-6 B B' p^2+ 2(B^2+B'^2) p^2 + 3(BB'p^2)^2 \right) \right\}  \, ,
 \end{eqnarray}
\item[3)] $J=1,J^{\prime}=0$
\begin{eqnarray}\label{eq:t10}
\overline{\sum} \sum |t|^2=\frac{1}{3} \,\overline{\sum} \sum |t|^2 (J=0,J^{\prime}=1)\, ,
 \end{eqnarray}
where the factor $\frac{1}{3}$ comes because we average over the initial $J=1$ polarizations.
\item[4)] $J=1,J^{\prime}=1$
\begin{eqnarray}\label{eq:t11}
\overline{\sum} \sum |t|^2 &=& \frac{1}{3} \,\frac{(A A')^2}{m_{\nu} m_l} \left\{ \frac{ 3\, m^2_l(M_{\rm inv}^{2(\nu l)}-m^2_l)}{M_{\rm inv}^{2(\nu l)}}  \left[(1+BB'p^2)^2+ \frac{2}{3}(B+B^{\prime})^2 p^2 \right] \right. \\ \nonumber \,
&+& \left. 2 \,\left(\widetilde{E}_\nu \widetilde{E}_l + \frac{1}{3}\widetilde{p}^2_\nu \right) \left[6 + 7(B^2+ B'^2)p^2 - 4 B B' p^2+6 (BB'p^2)^2 \right] \right\} .~~~~~~\,
 \end{eqnarray}
\end{itemize}
\end{itemize}
These techniques have also been used successfully in the evaluation of weak decays $M_1 \to M_2  M_3$ \cite{liangoset} and in $\tau^- \to M_1 M_2$ decays \cite{daioset}.

\section{Results}
The invariant mass distribution $d \Gamma/d M_{\rm inv}^{(\nu l)}$  is given for  $B \to \bar{\nu} l D$  by
\begin{eqnarray}
\frac{d \Gamma}{d M_{\rm inv}^{(\nu l)}}= \frac{2m_{\nu} 2m_l}{(2\pi)^3} \,\frac{1}{4M^2_B} p_D \widetilde{p}_\nu  \overline{\sum}\sum |t|^2  \, ,
\end{eqnarray}
where $p_D$ is the $D$ momentum in the $B$ rest frame and $\widetilde{p}_\nu$ the  $\bar\nu$ momentum in the $\nu l$ rest frame,
\begin{eqnarray}
p_D=\frac{\lambda^{1/2}(m^2_B,M_{\rm inv}^{2(\nu l)}, m^2_D)}{2 m_B}  \, .
\end{eqnarray}
By integrating $d \Gamma/d M_{\rm inv}^{(\nu l)}$  over $M_{\rm inv}^{(\nu l)}$ we obtain the width that we show in the tables.

\subsection{$B$  and $B^*$ decays}
We study only the most Cabibbo-favored processes, $b \to c$ and  $c \to s$. We show in Table \ref{tab:BPP} the $\bar{B}$, $\bar{B}_s$  and $\bar{B}_c$   semileptonic decays.
Since we can only provide ratios, we fix one decay rate to the experiment and the rest are predictions, In this case we fix our rate to ${B}^- \to D^0 ~e^- ~\bar{\nu}_{e}$ .
We then observe that the predictions done for six decays are all in agreement with experiment, except for the $\bar{B}^0 \to D^+ ~\tau^- \bar{\nu}_{\tau}$   that we will discuss later.
The $e^- ~\bar{\nu}_{e}$ and $\mu^- \bar{\nu}_{\mu}$  decay rates are very similar, since the masses of  $e^-$ and  $\mu^-$ are very small compared to the meson masses.
The term proportional  to $m^2_l$ in Eq. \eqref{eq:t00} is negligible for $e^-$ and $\mu^-$, but not for  $\tau^-$.
 This term is  responsible for a bigger rate than expected from phase space for the $\tau^- \bar{\nu}_{\tau}$ decays. In Table \ref{tab:BPP} we also show predictions for $\bar{B}^0_s$ and $B^-_c$ decays, for which
there are not yet experimental data in the PDG \cite{pdg}.
For the $\bar{B}^0 \to D^+ ~\tau^- \bar{\nu}_{\tau}$ we get a rate about a factor of two smaller than experiment. This has to be seen from the perspective  that we are implicitly  using the same form
factors as for $\bar{B}^0 \to D^+ ~e^- ~\bar{\nu}_{e}$. However, because of the larger $\tau^-$ mass, the momentum transfers are smaller in this latter case and by taking the same form factors as in
$\bar{B}^0 \to D^+ ~e^- ~\bar{\nu}_{e}$  we are reducing  the $\bar{B}^0 \to D^+ ~\tau^- \bar{\nu}_{\tau}$ rate more than one should.  This is telling us implicitly the strength  of the form factors
in the present reactions.  For ${B}^- \to D^0 ~\tau^- \bar{\nu}_{\tau}$ the experimental error is relatively large, such that the rate is compatible with the theoretical one, but also with double its value.

We would like to call  the attention to the rates for ${B}^- \to D^0 ~e^- ~\bar{\nu}_{e}$ (${B}^- \to D^0 ~\mu^- \bar{\nu}_{\mu}$). These rates are identical experimentally within
experimental errors,  and also theoretically (up to small difference due to the different masses of the mesons). This should be the case, since from one reaction to the other the only change
 has been to substitute  the $\bar{d}$ spectator quark in Fig. \ref{fig:diag} by a $\bar{u}$.


\begin{table}[h!]
\renewcommand\arraystretch{0.86}
\caption{Branching ratios for ${\color{blue}(PP)}$ semileptonic decay of $B$ meson.
We consider the same mean life $\tau$ for $\bar{B}^0$,${B}^-$, $\bar{B}^0_s$, but $\frac{\tau(B^-_c)}{\tau(B^-)}=0.31$}
\centering
\begin{tabular}{l  c  c  c }
\toprule[1.0pt]\toprule[1.0pt]
{Decay process}~~ ~& ~~~~~~ BR (Theo.) ~~~~~~   & ~~~~ BR (Exp.)\cite{pdg} ~~~\\
\hline
$\bar{B}^0 \to D^+ ~e^- ~\bar{\nu}_{e}$ & ~~$ 2.19\times 10^{-2}$~~  &~~$(2.19 \pm 0.12) \times 10^{-2}$~~ \\
$\bar{B}^0 \to D^+ ~\mu^- \bar{\nu}_{\mu}$ & ~~$2.17\times 10^{-2}$~~ &~~$(2.19 \pm 0.12) \times 10^{-2}$~~\\
$\bar{B}^0 \to D^+ ~\tau^- \bar{\nu}_{\tau}$ & ~~$ 4.97\times 10^{-3}$~~ &~~$(1.03 \pm 0.22) \times 10^{-2}$~~ \\
${B}^- \to D^0 ~e^- ~\bar{\nu}_{e}$ & ~{\color{red} fit the exp}  & ~~$(2.20 \pm 0.11) \times 10^{-2}$~~  \\
${B}^- \to D^0 ~\mu^- \bar{\nu}_{\mu}$ & ~~$2.19\times 10^{-2}$~~&~~$(2.20 \pm 0.11) \times 10^{-2}$~~   \\
${B}^- \to D^0 ~\tau^- \bar{\nu}_{\tau}$ & ~~$5.02\times 10^{-3}$~~ &~~$(7.7 \pm 2.5) \times 10^{-3}$~~ \\
$\bar{B}^0_s \to D^+_s ~e^- ~\bar{\nu}_{e}$ & ~~$ 2.07\times 10^{-2}$~~  &~~~~ \\
$\bar{B}^0_s \to D^+_s ~\mu^- \bar{\nu}_{\mu}$ & ~~$2.07\times 10^{-2}$~~ &~~~~  \\
$\bar{B}^0_s \to D^+_s~\tau^- \bar{\nu}_{\tau}$ & ~~$4.69\times 10^{-3}$~~ &~~~~  \\
$B^-_c \to \eta_c ~e^- ~\bar{\nu}_{e}$ & ~~$ 3.93\times 10^{-3}$~~  &~~~  \\
$B^-_c \to \eta_c ~\mu^- \bar{\nu}_{\mu}$ & ~~$3.90\times 10^{-3}$~~ &~~~~  \\
$B^-_c \to \eta_c ~\tau^- \bar{\nu}_{\tau}$ & ~~$8.49\times 10^{-4}$~~ &~~~~  \\
\bottomrule[1.0pt]\bottomrule[1.0pt]
\end{tabular}
\label{tab:BPP}
\end{table}

\begin{table}[h!]
\renewcommand\arraystretch{0.86}
\caption{The same as Table \ref{tab:BPP} but for  ${\color{blue}(PV)}$ case. }
\centering
\begin{tabular}{l  c  c  c }
\toprule[1.0pt]\toprule[1.0pt]
{Decay process}~~ ~& ~~~~~~ BR (Theo.) ~~~~~~   & ~~~~ BR (Exp.)\cite{pdg} ~~~ \\
\hline
$\bar{B}^0 \to D^{*+} ~e^- ~\bar{\nu}_{e}$ & ~~$ 4.86\times 10^{-2}$~~  &~~$(4.93 \pm 0.11) \times 10^{-2}$~~   \\
$\bar{B}^0 \to D^{*+} ~\mu^- \bar{\nu}_{\mu}$ & ~~$ 4.83\times 10^{-2}$~~ &$(4.93 \pm 0.11) \times 10^{-2}$  \\
$\bar{B}^0 \to D^{*+} ~\tau^- \bar{\nu}_{\tau}$ & ~~$1.03\times 10^{-2}$~~ &$(1.67 \pm 0.13) \times 10^{-2}$\\
${B}^- \to D^{*0} ~e^- ~\bar{\nu}_{e}$ & ~~$ 4.88\times 10^{-2}$~~  &$(5.69 \pm 0.19) \times 10^{-2}$ \\
${B}^- \to D^{*0} ~\mu^- \bar{\nu}_{\mu}$ & ~~$ 4.86 \times 10^{-2}$~~ &$(5.69 \pm 0.19) \times 10^{-2}$\\
${B}^- \to D^{*0} ~\tau^- \bar{\nu}_{\tau}$ & ~~$1.04\times 10^{-2}$~~ &$(1.88 \pm 0.2) \times 10^{-2}$  \\
$\bar{B}^0_s \to D^{*+}_s ~e^- ~\bar{\nu}_{e}$ & ~~$ 4.60\times 10^{-2}$~~  &~~~~ \\
$\bar{B}^0_s \to D^{*+}_s ~\mu^- \bar{\nu}_{\mu}$ & ~~$ 4.58\times 10^{-2}$~~ &~~~ \\
$\bar{B}^0_s \to D^{*+}_s ~\tau^- \bar{\nu}_{\tau}$ & ~~$ 9.64\times 10^{-3}$~~ &~~~~  \\
$B^-_c \to J/\psi ~e^- ~\bar{\nu}_{e}$ & ~~$ 9.47\times 10^{-3}$~~  &~~~~  \\
$B^-_c  \to J/\psi ~\mu^- \bar{\nu}_{\mu}$ & ~~$ 9.41\times 10^{-3}$~~ &~~~~  \\
$B^-_c \to J/\psi ~\tau^- \bar{\nu}_{\tau}$ & ~~$ 1.88\times 10^{-3}$~~ &~~~  \\
\bottomrule[1.0pt]\bottomrule[1.0pt]
\end{tabular}
\label{tab:BPV}
\end{table}

 In Table \ref{tab:BPV} we show results for $\bar{B}$ decays into $D^*$  and related reactions. This corresponds  to the case  $J=0, J'=1$ studied  in the former sections. We do not fit now one rate,
 because the idea is to make a prediction for these decays based on the $B \to D$ reactions. We  can see that the predicted rates for the case of light leptons are compatible with experiment,
 This should be seen as an accomplishment  of the present framework, which shows that assuming  the same form factors for $D$ or $D^*$  decay, as we would induce from heavy quark symmetry \cite{nieves},
 the rates for these two decays are a consequence  of the angular momentum  structure with the dynamics of the weak interaction.

 We can also observe that the  branching ratio for $\bar{B}^{0} \to D^{*+} ~\tau^- \bar{\nu}_{\tau}$ is  about  a factor $0.61$ the experimental one,  in line with what was observed in Table \ref{tab:BPP}.
 The same happens for ${B}^- \to D^{*0} ~\tau^- \bar{\nu}_{\tau}$ where the reduction factor is about $0.55$. Yet, if we evaluate the ratio of the rates of $\bar{B}^{0} \to D^{*+} ~\tau^- \bar{\nu}_{\tau}$
to $\bar{B}^{0} \to D^{+} ~\tau^- \bar{\nu}_{\tau}$ we find a factor of $2.07$ against  $1.62 \pm 0.37$ experimentally, or $2.07$ for the ratio of rates of
 ${B}^{-} \to D^{*0} ~\tau^- \bar{\nu}_{\tau}$ to $\bar{B}^{-} \to D^{0} ~\tau^- \bar{\nu}_{\tau}$  versus $2.44 \pm 0.83$  experimentally. One expects these two ratios  to be the same, and so they are
 experimentally  within errors, and  also compatible  with the theory.

 Once again we make predictions for six more decay modes. It is interesting to observe that the rates for ${B}^- \to \bar{D}^{*} \bar{\nu}_ l$
 are bigger experimentally than those of ${B}^- \to \bar{D} \bar{\nu}_ l$, something that is also obtained theoretically.

 Recently  much work has been devoted to the ratios
\begin{eqnarray}\label{eq:RD}
R_D&=&\frac{BR({B}^- \to D \bar{\nu}_{\tau} \tau )}{BR({B}^- \to D \bar{\nu}_{l} l )}=0.407 \pm 0.039 \pm 0.024 \nonumber \,,\\
R_{D^*}&=&\frac{BR({B}^- \to D^* \bar{\nu}_{\tau} \tau )}{BR({B}^- \to D^* \bar{\nu}_{l} l )}=0.304 \pm 0.013 \pm 0.007  \, ,
\end{eqnarray}
 from where one expects to observe new physics \cite{Fajfer,German}. The values in Eq. \eqref{eq:RD}  are taken from the HFLAV collaboration average \cite{HFLAV}.
The most precise single measurement performed so far is the recent one of the LHCb Collaboration \cite{lhcbd}  $R_{D^*}=0.291\pm 0.019 \pm 0.026\pm 0.013$.
Our result for these ratios are $R_D=0.23$,  $R_{D^*}=0.211$. As we can see, both $R_D$ and $R_{D^*}$ are smaller than experiment for the reasons discussed above.
To put our  results  in perspective  we can compare   these result with Lattice results \cite{lat1,lat2} which give $R_D=0.299\pm 0.011$, and calculations based on
the standard model obtaining ratios of form factors from experiment \cite{Fajfer} which give $R_{D^*}=0.252 \pm 0.03$. This latter case is increased to $0.27$ in \cite{Genaro}
by taking into account in the theoretical evaluation that $D^* \to D\pi$, the mode where the $D^*$ is observed experimentally.

Another ratio of interest  is
\begin{eqnarray}\label{eq:Jpsi}
R_{J/\psi} = \frac{BR(B^+_c \to J/\psi ~ {\nu}_{\tau}\tau^+)}{BR(B^+_c \to J/\psi ~ {\nu}_{\mu} \mu^+)} =0.71  \pm 0.12 \pm 0.18 \, ,
\end{eqnarray}
reported by LHCb \cite{lhcbj}. We obtain $0.20$ for this ratio, short of the experimental one even considering errors, and one must find the reason in the discussion in former points since
momentum transfer for $B^+_c \to J/\psi ~\tau^+ {\nu}_{\tau}$ is smaller than in $B^+_c  \to J/\psi ~\mu^+ {\nu}_{\mu}$.
 It is also useful to compare our results with
other theoretical works  based on  the standard model which provide value around  $0.25 \sim 0.28$ \cite{Semay,Kiselev,Ivanov,eli} for this ratio.

There is another remark we can do in view of our easy expressions for
$\overline{\sum} \sum |t|^2$.  If one looks  at Eq. \eqref{eq:t00}  for  $J=0,J^{\prime}=0$,  one can see that the term independent of $p$  is proportional  to $m^2_l$, which is very small for light leptons.
The important term is this case is the second term of that equation,  proportional  to $p^2$. This tells us that nonrelativistic calculations, which would neglect this momentum,
or the strict use of heavy quark symmetry, neglecting terms of $O(\frac{p}{m_Q})$ (see $B$ factor  in Eq. \eqref{eq:wfn}),
would provide very  bad results
for the rate of $\bar{B} \to D$  and light leptons.

\begin{table}[h!]
\renewcommand\arraystretch{0.86}
\caption{Widths for ${\color{blue}(VP)}$ semileptonic decay of $B^*$ mesons in units of $\Gamma(B^- \to {D}^0  e^- ~\bar{\nu}_{e})$. }
\centering
\begin{tabular}{l  c  c  c }
\toprule[1.0pt]\toprule[1.0pt]
{Decay process} ~~~~& ~~~~  $\Gamma$ (Theo.) ~~~~ & ~~~~$\Gamma$ (Ref. \cite{wangwang})  ~~~~&  ~~~~$\Gamma$ (Ref. \cite{changyang}) ~~~~\\
\hline
$\bar{B}^{*0} \to D^+ ~e^- ~\bar{\nu}_{e}$ & 0.95  & \\
$\bar{B}^{*0} \to D^+ ~\mu^- \bar{\nu}_{\mu}$ & 0.95 &  \\
$\bar{B}^{*0} \to D^+ ~\tau^- \bar{\nu}_{\tau}$ & 0.24& \\
${B}^{*-} \to D^0 ~e^- ~\bar{\nu}_{e}$ & 0.96 & 1.74& 1.21\\
${B}^{*-} \to D^0 ~\mu^- \bar{\nu}_{\mu}$ & 0.96 & 1.73& 1.21 \\
${B}^{*-} \to D^0 ~\tau^- \bar{\nu}_{\tau}$ & 0.24 & 0.43& 0.36 \\
$\bar{B}^{*0}_s \to D^+_s ~e^- ~\bar{\nu}_{e}$ &0.91 & 1.57 & 1.07  \\
$\bar{B}^{*0}_s \to D^+_s ~\mu^- \bar{\nu}_{\mu}$ & 0.91 & 1.56 & 1.07 \\
$\bar{B}^{*0}_s \to D^+_s ~\tau^- \bar{\nu}_{\tau}$ & 0.23& 0.41 & 0.31 \\
${B^{*-}_c} \to \eta_c ~e^- ~\bar{\nu}_{e}$ & 0.59 & 1.09&\\
${B^{*-}_c}\to \eta_c ~\mu^- \bar{\nu}_{\mu}$ & 0.58& 1.09&\\
${B^{*-}_c} \to \eta_c ~\tau^- \bar{\nu}_{\tau}$ & 0.14 &  0.33&\\
\bottomrule[1.0pt]\bottomrule[1.0pt]
\end{tabular}
\label{tab:BVP}
\end{table}

In Table \ref{tab:BVP} we also show the rate for the semileptonic decays of $\bar{B}^* \to D$. These decay rates have not been observed
and one reason maybe the fact that $B^*$ decays electromagnetically in $\gamma B$. In Ref. \cite{wangwang} these rates were evaluated  using the
Bethe-Salpeter approach with the instantaneous approximation. We compare our predictions with those in Ref. \cite{{wangwang}}. These decay widths are also
evaluated in Ref. \cite{changyang} in the Baner-Stech-Wirbel  model and in  Table \ref{tab:BVP} we also show these results. Our results are qualitatively similar  to those
of \cite{changyang}, about $25\%$  smaller  and also smaller than those of \cite{wangwang} by a factor of about $0.6$. One should stress the simplicity of our approach with respect
to \cite{wangwang} where 14 form factors  are evaluated and the theory  relies on  several parameters  \cite{zhangwang}, partly constrained from the masses of the mesons.
We should stress  that the rates for $B^* \to D \bar{\nu} l$ in our approach  involve  the same matrix element as for $B \to D^* \bar{\nu} l$
 ($J=1,J^{\prime}=0$, versus $J=0,J^{\prime}=1$ in Eqs.  \eqref{eq:t01},\eqref{eq:t10}). Given the accuracy by which we predicted the rate for  $B \to D^* \bar{\nu} l$
 in Table \ref{tab:BPV}, our predicted rates for $B^* \to D \bar{\nu} l$ should be equally accurate.  Yet, seen from the perspective that these are the first theoretical predictions for
 $B^*$  decay rates, the main message is that the three calculations reported provide very similar numbers and this should be sufficient for planning possible experimental searches.

We complete this part by making predictions for $B^* \to D^* \nu l$ in Table \ref{tab:BVV}, in which we also compare our results
with those obtained in Ref. \cite{wangwang}.  We observe now that in both cases the widths are much bigger than for $B^* \to D \nu l$, and
our predictions are again a factor of about $0.55$  those  of Ref. \cite{wangwang}, with a bit bigger discrepancies for decays in the $\tau$ mode, as we would expect.

\begin{table}[h!]
\renewcommand\arraystretch{0.86}
\caption{The same as Table \ref{tab:BVP} but for  ${\color{blue}(VV)}$ case.  }
\centering
\begin{tabular}{l  c  c   }
\toprule[1.0pt]\toprule[1.0pt]
{Decay process}~~ ~& ~~~~~~ $\Gamma$ (Theo.) ~~~~~~   & ~~~~ $\Gamma$ (Ref. \cite{wangwang})  ~~~ \\
\hline
$\bar{B}^{*0} \to D^{*+} ~e^- ~\bar{\nu}_{e}$ & 2.52  &  \\
$\bar{B}^{*0} \to D^{*+} ~\mu^- \bar{\nu}_{\mu}$ & 2.50 & \\
$\bar{B}^{*0} \to D^{*+} ~\tau^- \bar{\nu}_{\tau}$ & 0.51 & \\
${B}^{*-} \to D^{*0} ~e^- ~\bar{\nu}_{e}$ & 2.53 & 4.98 \\
${B}^{*-} \to D^{*0} ~\mu^- \bar{\nu}_{\mu}$ & 2.51 & 4.95 \\
${B}^{*-} \to D^{*0} ~\tau^- \bar{\nu}_{\tau}$ & 0.51 &1.08 \\
$\bar{B}^{*0}_s \to D^{*+}_s ~e^- ~\bar{\nu}_{e}$ & 2.40&4.40 \\
$\bar{B}^{*0}_s \to D^{*+}_s ~\mu^- \bar{\nu}_{\mu}$ & 2.37  & 4.38  \\
$\bar{B}^{*0}_s \to D^{*+}_s ~\tau^- \bar{\nu}_{\tau}$ & 0.48 &1.00  \\
${B^{*-}_c} \to J/\psi ~e^- ~\bar{\nu}_{e}$\footnote{Here we take the value $m_{B^{*-}_c} = 6.333$ GeV predicted by the quark potential model \cite{ebert11}.}  & 1.60 &2.95 \\
${B^{*-}_c}\to J/\psi ~\mu^- \bar{\nu}_{\mu}$ & 1.58 & 2.94 \\
${B^{*-}_c}\to J/\psi ~\tau^- \bar{\nu}_{\tau}$ & 0.31& 0.82  \\
\bottomrule[1.0pt]\bottomrule[1.0pt]
\end{tabular}
\label{tab:BVV}
\end{table}

\begin{figure}[ht]
\includegraphics[scale=1.1]{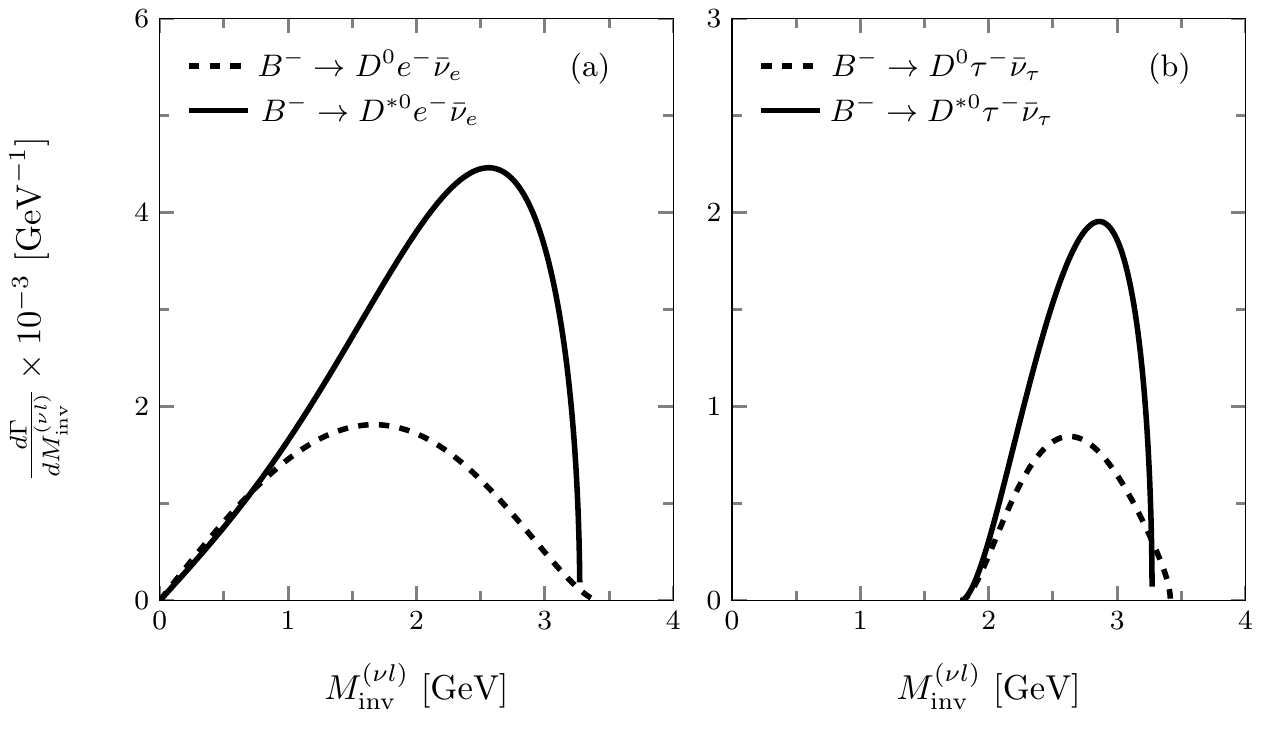}
\caption{The differential width of  semileptonic decay of $B$  meson decay.}
\label{fig:BP}
\end{figure}

\begin{figure}[ht]
\includegraphics[scale=1.1]{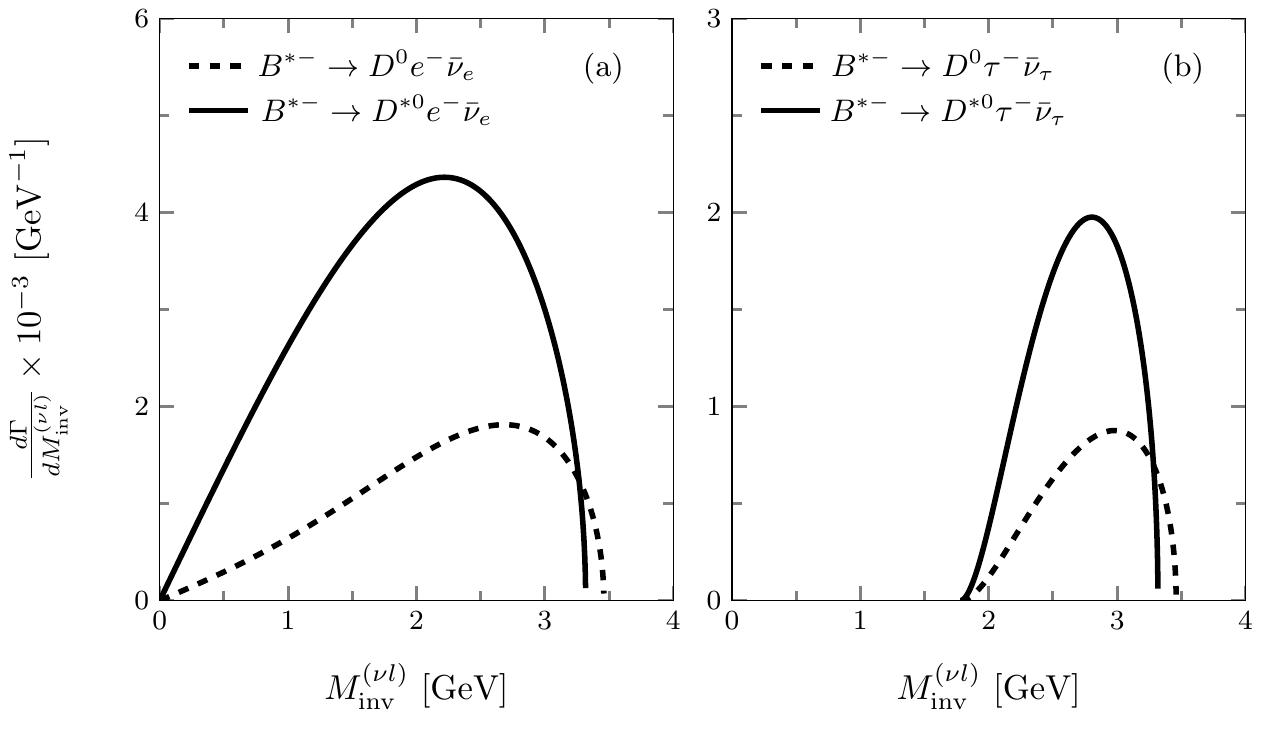}
\caption{The differential width of  semileptonic decay of  $B^{*}$ meson decay.}
\label{fig:BV}
\end{figure}

It is also interesting to look into the  invariant mass distribution $d \Gamma/d M_{\rm inv}^{(\nu l)}$.  In Fig. \ref{fig:BP} we show $d \Gamma/d M_{\rm inv}^{(\nu l)}$
for ${B}^- \to D^0 ~e^- ~\bar{\nu}_{e}$, ${B}^- \to D^{*0} ~e^- ~\bar{\nu}_{e}$,  and  ${B}^- \to D^0 ~\tau^- \bar{\nu}_{\tau}$, ${B}^- \to D^{*0} ~\tau^- \bar{\nu}_{\tau}$.
In the case of ${B}^- \to D^0 ~e^- ~\bar{\nu}_{e}$ the mass distribution peaks around $1.7~\rm{GeV}$ while for ${B}^- \to D^{*0} ~e^- ~\bar{\nu}_{e}$ it peaks around $2.7~\rm{GeV}$.
This difference  is surprising  in view of the small  difference  of mass between $D$ and $D^*$. One must see the reason  for this behaviour  in the structure of Eqs. \eqref{eq:t00}
and \eqref{eq:t01}.  As we mentioned before, the ${B}^- \to D^0 ~e^- ~\bar{\nu}_{e}$ reaction gets most of its strength from the $p^2$ term of Eq. \eqref{eq:t00}  since the $m^2_l$  term
is extremely small.
One gets a bigger  contribution  the larger  $p^2$, but  this means a smaller $M_{\rm inv}^{(\nu l)}$(see Eq. \eqref{eq:p}). In the case of  ${B}^- \to D^{*0} ~e^- ~\bar{\nu}_{e}$,
Eq.\eqref{eq:t01}  there are large terms  independent of $p$ and the argument does not hold. In the case of  ${B}^- \to D^0 ~\tau^- \bar{\nu}_{\tau}$ and  ${B}^- \to D^{*0} ~\tau^- \bar{\nu}_{\tau}$,
the $m^2_l$ term is not so small and the position of the peaks is much closer. The argument discussed above become even more clear when we look at the distributions of Fig. \ref{fig:BV} for
${B}^{*-} \to D^0 ~e^- ~\bar{\nu}_{e}$, ${B}^{*-} \to D^{*0} ~e^- ~\bar{\nu}_{e}$ and ${B}^{*-} \to D^0 ~\tau^- \bar{\nu}_{\tau}$, ${B}^{*-} \to D^{*0} ~\tau^- \bar{\nu}_{\tau}$.  In this case
$\overline{\sum} \sum |t|^2$ is given by Eqs. \eqref{eq:t01} and \eqref{eq:t11} and both expression are sizeable in the limit of $p\rightarrow 0$, as a consequence  of which, the shapes
of the mass distributions are now very different than in the former case.

\subsection{$D$  and $D^*$ decays}
In Table \ref{tab:DPP}  we show results for $D^+ \to \bar{K}^0 e^+~{\nu}_{e}$ and related reactions. We fix our results to the rate
for $D^0 \to {K}^-  e^+ ~{\nu}_{e}$. Our results are in fair  agreement for other reactions.




\begin{table}[h!]
\renewcommand\arraystretch{0.86}
\caption{Branching ratios for ${\color{blue}(PP)}$ semileptonic decay of $D$ meson. We have taken into account
that $\frac{\tau_{D^+}}{\tau_{D^0}}=2.54$ and $\frac{\tau_{D_s^+}}{\tau_{D^0}}=1.23$.  For $\eta$ and $\eta^{\prime}$ production we consider that the weights for the $s\bar{s}$
 components are $\frac{1}{3}$ and $\frac{2}{3}$, respectively.}
\centering
\begin{tabular}{l  c  c   }
\toprule[1.0pt]\toprule[1.0pt]
{Decay process}~~ ~& ~~~~~~ BR (Theo.) ~~~~~~   & ~~~~ BR (Exp.) \cite{pdg}~~~ \\
\hline
$D^+ \to \bar{K}^0 e^+~{\nu}_{e}$ & ~~$ 8.94\times 10^{-2}$~~  &~~$(8.82 \pm 0.13) \times 10^{-2}$~~ \\
$D^+ \to \bar{K}^0 \mu^+ {\nu}_{\mu}$ & ~~$8.61 \times 10^{-2}$~~ &~~$(8.74 \pm 0.19) \times 10^{-2}$~~ \\
$D^0 \to {K}^-  e^+ ~{\nu}_{e}$ & ~{\color{red} fit the exp}  &~~$(3.53 \pm 0.028) \times 10^{-2}$~~  \\
$D^0 \to {K}^- \mu^+ {\nu}_{\mu}$ & ~~$3.40\times 10^{-2}$~~ &~~$(3.31 \pm 0.13) \times 10^{-2}$~~ \\
$D^+_s \to \eta e^+ ~{\nu}_{e}$ & ~~$ 1.53\times 10^{-2}$~~  &~~$(2.29 \pm 0.19) \times 10^{-2}$~ \\
$D^+_s \to \eta \mu^+ {\nu}_{\mu}$ & ~~$1.48\times 10^{-2}$~~  \\
$D^+_s \to \eta' e^+ ~{\nu}_{e}$ & ~~$ 5.16\times 10^{-3}$~~  &~~$(7.4 \pm 1.4) \times 10^{-3}$ \\
$D^+_s  \to \eta' \mu^+ {\nu}_{\mu}$ & ~~$ 4.87 \times 10^{-3}$~~ & ~~ \\
\bottomrule[1.0pt]\bottomrule[1.0pt]
\end{tabular}
\label{tab:DPP}
\end{table}

\begin{table}[h!]
\renewcommand\arraystretch{0.86}
\caption{The same as Table \ref{tab:DPP} but for  ${\color{blue}(PV)}$ case. }
\centering
\begin{tabular}{l  c  l   }
\toprule[1.0pt]\toprule[1.0pt]
{Decay process }~~ ~& ~~~~~~ BR (Theo.) ~~~~~~   & ~~~~ BR (Exp.) \cite{pdg}~~~ \\
\hline
$D^+ \to \bar{K}^{*0}~e^+ {\nu}_{e}$ & ~~$ 4.21\times 10^{-2}$~~  &$\frac{3}{2}\times(3.66 \pm 0.12) \times 10^{-2}$ \\
$D^+ \to \bar{K}^{*0} ~\mu^+ ~{\nu}_{\mu}$ & ~~$3.96\times 10^{-2}$~~ &$\frac{3}{2}\times(3.52 \pm 0.10) \times 10^{-2}$ \\
$D^0 \to {K}^{*-} ~ e^+ ~{\nu}_{e}$ & ~~$ 1.66\times 10^{-2}$~~  &$(2.15 \pm 0.16) \times 10^{-2}$ \\
$D^0 \to {K}^{*-} ~\mu^+ ~{\nu}_{\mu}$ & ~~$1.56\times 10^{-2}$~~ &$(1.86 \pm 0.24) \times 10^{-2}$ \\
$D^+_s\to \phi~e^+ ~{\nu}_{e}$ & ~~$ 1.63\times 10^{-2}$~~  &$(2.39 \pm 0.23) \times 10^{-2}$~~  \\
$D^+_s  \to \phi~ \mu^+ {\nu}_{\mu}$ & ~~$1.54\times 10^{-2}$~~ &  \\
\bottomrule[1.0pt]\bottomrule[1.0pt]
\end{tabular}
\label{tab:DPV}
\end{table}

In Table \ref{tab:DPV} we show the results for $D^+ \to \bar{K}^{*0}~e^+ {\nu}_{e}$ reaction and related ones.
The agreement with experiment  is fair. In particular if we look at ratios for different final vectors we find
 \begin{eqnarray}\label{eq:DDs}
 \frac{BR( D^0 \to {K}^{*-} ~ e^+ ~{\nu}_{e})}{BR(D^+_s\to \phi~e^+ ~{\nu}_{e})}=1.02 \, ,
\end{eqnarray}
when experimentally is $0.9 \pm 0.11$, which  is compatible with Eq. \eqref{eq:DDs}.

\begin{table}[h!]
\renewcommand\arraystretch{0.86}
\caption{Widths for ${\color{blue}(VP)}$ semileptonic decay of $D^*$ mesons in units of $\Gamma(D^0 \to {K}^-  e^+ ~{\nu}_{e})$. }
\centering
\begin{tabular}{l  c  c  }
\toprule[1.0pt]\toprule[1.0pt]
{Decay process}~~ ~& ~~~~~~ $\Gamma$ (Theo.) ~~~~~~   & ~~~~ $\Gamma$ (Exp.) ~~~ \\
\hline
$D^{*+} \to \bar{K}^0~e^+ {\nu}_{e}$ & 1.25  &~~ \\
$D^{*+}  \to \bar{K}^0 \mu^+ {\nu}_{\mu}$ &1.21  &~~  \\
$D^{*0}  \to {K}^-  e^+ {\nu}_{e}$ & 1.25  &~~~~ \\
$D^{*0} \to {K}^- \mu^+ {\nu}_{\mu}$ & 1.22 &~~~ \\
$D^{*+}_s \to \eta e^+ {\nu}_{e}$ & 0.44  &~~~\\
$D^{*+}_s   \to \eta \mu^+ {\nu}_{\mu}$ & 0.43 & ~  \\
$D^{*+}_s  \to \eta'  e^+ {\nu}_{e}$ & 0.19 &~~~~ \\
$D^{*+}_s   \to \eta' \mu^+ {\nu}_{\mu}$ & 0.18& ~~  \\
\bottomrule[1.0pt]\bottomrule[1.0pt]
\end{tabular}
\label{tab:DVP}
\end{table}

\begin{table}[h!]
\renewcommand\arraystretch{0.86}
\caption{The same as Table \ref{tab:DVP} but for  ${\color{blue}(VV)}$ case.  }
\centering
\begin{tabular}{l  c  c   }
\toprule[1.0pt]\toprule[1.0pt]
{Decay process }~~ ~& ~~~~~~ $\Gamma$ (Theo.) ~~~~~~   & ~~~~ $\Gamma$ (Exp.) ~~~\\
\hline
$D^{*+} \to \bar{K}^{*0}~e^+ {\nu}_{e}$ & 0.85  &  \\
$D^{*+} \to \bar{K}^0 \mu^+ {\nu}_{\mu}$ & 0.80 &  \\
$D^{*0} \to {K}^{*-}  e^+ {\nu}_{e}$ &  0.86 &  \\
$D^{*0} \to {K}^{*-} \mu^+ {\nu}_{\mu}$ & 0.80 &  \\
$D^{*+}_s  \to \phi~e^+ ~{\nu}_{e}$ & 0.70 &\\
$D^{*+}_s  \to \phi~ \mu^+ {\nu}_{\mu}$ & 0.66 &   \\
\bottomrule[1.0pt]\bottomrule[1.0pt]
\end{tabular}
\label{tab:DVV}
\end{table}

Finally, in  Table \ref{tab:DVP} and \ref{tab:DVV} we give for completeness the rate for $D^{*+} \to \bar{K}^0~e^+ {\nu}_{e}$ and $D^{*0}  \to {K}^-  e^+ {\nu}_{e}$
and related reactions.  Yet, the fact that the $D^*$  decays strongly  into $D \pi$, makes the observation of these modes  extremely difficult and we do not elaborate
further on them.

\begin{figure}[ht!]
\includegraphics[scale=1.1]{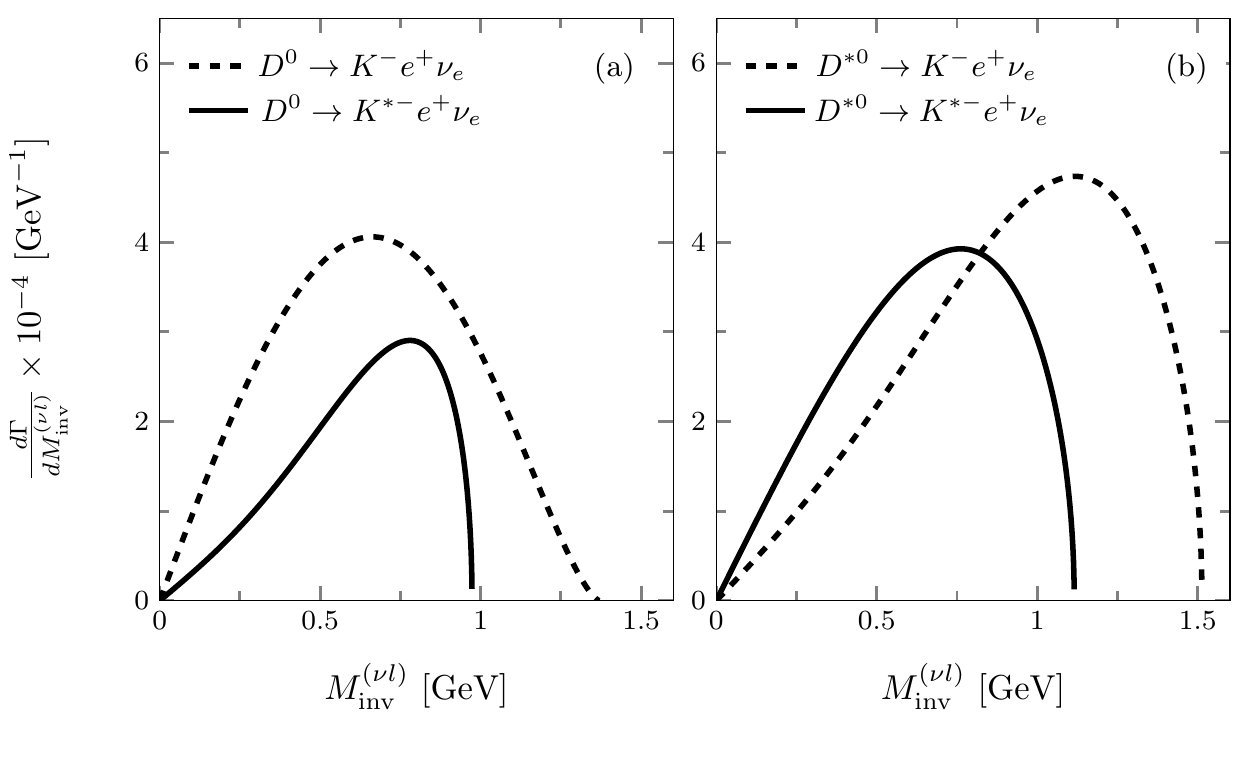}
\caption{The differential width of  semileptonic decay of $D$ or $D^{*}$ meson decay.}
\label{fig:DPV}
\end{figure}

To finish the section we show  in Fig. \ref{fig:DPV} the mass distribution for $D^{0}  \to {K}^-  e^+ {\nu}_{e}$, $D^{0} \to {K}^{*-}  e^+ {\nu}_{e}$
 and  $D^{*0}  \to {K}^-  e^+ {\nu}_{e}$, $D^{*0} \to {K}^{*-}  e^+ {\nu}_{e}$. We can see that the arguments discussed earlier about the relative peak  positions
 of these  kind of reactions $J=0, J'=0$ versus $J=0, J'=1$  and $J=1, J'=0$ versus $J=1, J'=1$ also hold in this case, although the changes are not so drastic
 as in Fig.~\ref{fig:BP} because of the restrictions on the phase space for $D^{0}  \to {K}^{*-}  e^+ {\nu}_{e}$.

\section{Conclusions}
We have performed the angular momentum algebra in the evaluation of the weak matrix element in the semileptonic decays of $B^{(*)}$, $D^{(*)}$ mesons.
The matrix elements of the weak current  are evaluated  in the rest frame of the $\nu l$ pair, which require spinors with finite (and large) momentum,  and we obtain
finally very simple expressions for the matrix elements involved and the sum over polarizations  of their modulus squared. In terms of these analytical expressions
we can give easy explanations for peculiarities in the invariant mass distributions  and the ratios of rates for different reactions. By fixing the normalization
of the theoretical rates to the experimental one of a reaction in the $B$ sector and another one in the $D$ sector, we can obtain the rate for the rest
of the reactions. The agreement with experiment is good, with   discrepancies  for the production  of $\tau$ leptons which we can trace to the different momentum transfers involved
in the production  of light $e$,$\mu$ leptons and $\tau$ lepton.

 One of the output of the study is the prediction of decay rates for $B^*$ and  $D^*$, which have been the object of discussion  recently since they could be observed in future measurements of the LHCb
collaboration. We justify that our predictions for these decay widths should be very accurate, which can be used in planning experiments to observe them  in the future.

\section*{Acknowledgments}
We wish to express our  thanks to Juan Nieves for his valuable help.  L. R. D. acknowledges the support from the National
Natural Science Foundation of China (No. 11575076) and the State Scholarship Fund of China (No. 201708210057).
This work is partly supported by the Spanish Ministerio de Economia y Competitividad
and European FEDER funds under Contracts No. FIS2017-84038-C2-1-P B
and No. FIS2017-84038-C2-2-P B, and the Generalitat Valenciana in the program Prometeo II-2014/068, and
the project Severo Ochoa of IFIC, SEV-2014-0398.

\newpage
\appendix
\section{Evaluation of the matrix elements}
\subsection{Evaluation of $M_0$}
We start from the expression of $M_0$ in Eq.~\eqref{eq:g0}
\begin{eqnarray}
M_0= A A'  \delta_{s s''}  \left\{1+B B^{\prime}  p^2 - (B+B^{\prime})\,p \,\sqrt{3}\, {\cal C}(\frac{1}{2} 1 \frac{1}{2}; s,0,s)\right\} \,.
\end{eqnarray}
By looking at the spin third components of Fig. \ref{fig:diag}, we must combine $s,s'$ to give $J M$ and $s'',s'$ to give $J' M'$.  Thus we have
\begin{eqnarray}\label{eq:m0}
\sum_{s,s',s''} {\cal C}\left(\frac{1}{2} \frac{1}{2} J; s,s',M\right) {\cal C}\left(\frac{1}{2} \frac{1}{2} J'; s'',s',M'\right) AA' \left\{1+B B^{\prime} p^2 -(B+B^{\prime})p\sqrt{3}\,{\cal C}(\frac{1}{2} 1 \frac{1}{2}; s,0,s)\right\} \delta_{s s''} \,. ~~~~~~~
\end{eqnarray}
From here we conclude that $s+s'=M$ and $s''+s'=M'$, and since $s''=s$ then $M=M'$.

For the first two terms in Eq.~\eqref{eq:m0} we get
\begin{eqnarray}
\sum_{s} {\cal C}\left(\frac{1}{2} \frac{1}{2} J; s,M-s,M\right) {\cal C}\left(\frac{1}{2} \frac{1}{2} J'; s,M-s,M\right)=\delta_{JJ'} \,.
\end{eqnarray}

For the third term, we have a combination of three Clebsch-Gordan coefficients (CGC).  We follow the angular momentum algebra of Rose \cite{rose} and write
\begin{eqnarray}
{\cal C}\left(\frac{1}{2} \frac{1}{2} J; s,M-s,M \right) &=& (-1)^{\frac{1}{2}-s} \sqrt{\frac{2J+1}{2}} \  {\cal C}\left(J \frac{1}{2} \frac{1}{2}; M,-s,M-s \right) , \nonumber\\
{\cal C}\left(\frac{1}{2} \frac{1}{2} J'; s,M-s,M \right)& =& (-1)^{\frac{1}{2}+\frac{1}{2}-J'}  \  {\cal C}\left(\frac{1}{2} \frac{1}{2} J'; M-s,s, M\right) , \nonumber\\
{\cal C}\left(\frac{1}{2} 1 \frac{1}{2}; s,0,s \right) &=& (-1)^{\frac{1}{2}-s} \sqrt{\frac{2}{3}} \  {\cal C}\left(\frac{1}{2} \frac{1}{2} 1; s,-s,0\right) ,
\end{eqnarray}
and then
\begin{eqnarray}
&&\sum_s \mathcal{C}\left(\frac{1}{2} \frac{1}{2} J; s,M-s,M \right) \mathcal{C}\left(\frac{1}{2} \frac{1}{2} J'; s,M-s,M \right) \sqrt{3}\ \mathcal{C}\left(\frac{1}{2} 1 \frac{1}{2}; s,0,s \right) \nonumber\\
&=&(-1)^{1-J'} \sqrt{2J+1} \ \mathcal{C}\left(J \frac{1}{2} \frac{1}{2}; M,-s,M-s \right)  \mathcal{C}\left(\frac{1}{2} \frac{1}{2} J'; M-s,s, M\right)
 \mathcal{C}\left(\frac{1}{2} \frac{1}{2} 1; s,-s,0\right)\nonumber\\
&=&(-1)^{1-J'} \sqrt{2J+1}\sqrt{6}  \ \mathcal{W}\left(J \frac{1}{2} J' \frac{1}{2}; \frac{1}{2} 1 \right)\mathcal{C}\left(J 1 J'; M,0, M\right) \,,
\end{eqnarray}
where $\mathcal{W}(\cdots)$ is a  Racah coefficient.

Altogether
\begin{eqnarray}\label{eq:t11}
M_0 &=& A A' \bigg{\{}  (1+BB'p^2) \delta_{M M'} \,  \delta_{J J'} \nonumber\\
&-& \left . (B+B')\ p  \, \delta_{M M'} \sqrt{6} \, \sqrt{2J+1} (-1)^{1-J'} \mathcal{W}\left(J \frac{1}{2} J' \frac{1}{2}; \frac{1}{2} 1 \right)\mathcal{C}\left(J 1 J'; M,0, M\right)\right\} \,.
 \end{eqnarray}
Calculating explicitly the Racah coefficient and the CGC, we find
\begin{itemize}
 \item[1)] $J=0,J'=0$
\begin{eqnarray} \label{eq:M00}
M_0 =  A A'(1+B B'p^2) \,\delta_{M 0} \, \delta_{M' 0} \, .
\end{eqnarray}
 \item[2)] $J=0,J'=1$
  \begin{eqnarray} \label{eq:M01}
M_0 = - A A'(B+B') \, p \,\delta_{M 0} \,\delta_{M' 0} \, .
\end{eqnarray}
\item[3)] $J=1,J'=0$
  \begin{eqnarray} \label{eq:M10}
M_0 = - A A'(B + B') \,p \,\delta_{M 0} \, \delta_{M' 0} \, .
\end{eqnarray}
\item[4)] $J=1,J'=1$
  \begin{eqnarray} \label{eq:M11}
M_0 =  A A'\left\{(1+B B'p^2)\delta_{M M'} - \sqrt{2} \,\delta_{M M'}\,(B+B') \, p \,\mathcal{C}(1 1 1;M,0,M)\right\} \,.
\end{eqnarray}
\end{itemize}
\subsection{Evaluation of $N_{\mu}$}
As seen in  Eq. \eqref{eq:gi2}, we had $\gamma_i-\gamma_i\gamma_5$ written in spherical basis,
\begin{eqnarray} \label{eq:Nnu}
N_{\mu} =A A'
\left\{1-B\,p \, (-1)^{\frac{1}{2}-s} -B'\,p \, (-1)^{\frac{1}{2}-s''}
+ B B^{\prime} p^2 (-1)^{1-s-s''} \right\}  \langle s''| {{\sigma}}_\mu |s \rangle  \, ,
\end{eqnarray}
and now we must project over $J M$ and $J' M'$. Thus, we get the  combination
\begin{eqnarray} \label{eq:Nnua}
\sum_{s,s',s''} \mathcal{C}\left(\frac{1}{2} \frac{1}{2} J; s,s',M \right) \mathcal{C}\left(\frac{1}{2} \frac{1}{2} J'; s'',s',M' \right) \sqrt{3}\ \mathcal{C}\left(\frac{1}{2} 1 \frac{1}{2}; s,\mu,s'' \right)  \, ,
\end{eqnarray}
which implies $s+s'=M$ and $s''+s'=M'$, and then we get $s''+M-s=M'$, and $s+\mu=s''$,   $s+\mu+M-s=M'$, so $\mu=M'-M$.

Hence, we get the sum
\begin{eqnarray}\label{eq:S}
S=\sum_s \mathcal{C}\left(\frac{1}{2} \frac{1}{2} J; s,M-s,M \right) \mathcal{C}\left(\frac{1}{2} \frac{1}{2} J'; M'-M+s,M-s,M' \right)  \nonumber\\
\times \sqrt{3}\ \mathcal{C}\left(\frac{1}{2} 1 \frac{1}{2}; s,M'-M,M'-M+s \right) \, ,
\end{eqnarray}
which goes with the term $"1"$ in the bracket of Eq. \eqref{eq:Nnu}.

Permuting indices in the CGC, we have
\begin{eqnarray}
 \mathcal{C}\left(\frac{1}{2} \frac{1}{2} J; s,M-s,M \right) &=& (-1)^{\frac{1}{2}-s} \sqrt{\frac{2J+1}{2}} \  \mathcal{C}\left(J \frac{1}{2} \frac{1}{2}; M,-s,M-s \right) , \nonumber\\
 \mathcal{C}\left(\frac{1}{2} \frac{1}{2} J'; M'-M+s,M-s,M' \right)& =& (-1)^{1-J'}  \  \mathcal{C}\left(\frac{1}{2} \frac{1}{2} J'; M-s,M'-M+s, M'\right) , \nonumber\\
 \mathcal{C}\left(\frac{1}{2} 1 \frac{1}{2}; s,M'-M,M'-M+s \right) &=& (-1)^{\frac{1}{2}-s} \sqrt{\frac{2}{3}} \  \mathcal{C}\left(\frac{1}{2} \frac{1}{2} 1; s,M-M'-s,M-M'\right) , \nonumber\\
 &=&(-1)^{\frac{1}{2}-s} \sqrt{\frac{2}{3}} \mathcal{C}\left(\frac{1}{2} \frac{1}{2} 1; -s,M'-M+s,M'-M\right) \, .~~~~~~~~~
\end{eqnarray}
Then the expression $S$ of Eq.~\eqref{eq:S} becomes
\begin{eqnarray}
S&=& (-1)^{1-J'} \sqrt{2J+1} \sum_s \mathcal{C}\left(J \frac{1}{2} \frac{1}{2}; M, -s,M-s \right)  \nonumber\\
 &\times & \mathcal{C}\left(\frac{1}{2} \frac{1}{2} J'; M-s,M'-M+s,M'\right) \mathcal{C}\left(\frac{1}{2} \frac{1}{2} 1 ; -s,M'-M+s,M'-M \right)  \nonumber\\
 &=& (-1)^{1-J'} \sqrt{2J+1} \sqrt{6} \, \mathcal{W}\left(J \frac{1}{2} J' \frac{1}{2}; \frac{1}{2} 1 \right) \mathcal{C}\left(J 1 J'; M, M'-M,M'\right) \, .
\end{eqnarray}

The term $BB'p^2(-1)^{1-s-s''}$ in Eq.~\eqref{eq:Nnu} is easy, since $s+s''=M'-M$ and we get the term $1+BB'p^2(-1)^{M-M'}$
times the expression Eq.~\eqref{eq:Nnua}.

Next we must evaluate the other two terms in Eq.~\eqref{eq:Nnu} that have an extra phase,
\begin{eqnarray}
(-1)^{\frac{1}{2}-s} \,;\qquad  (-1)^{\frac{1}{2}-s''}=(-1)^{\frac{1}{2}+M-M'-s}=(-1)^{\frac{1}{2}-s}  \, (-1)^{M-M'}
\end{eqnarray}
Since $(-1)^{M-M'}$ is not affected by the $s$ sum, we only have one structure to calculate.  Incorporating an extra phase in the sum is not trivial.
To finally be able to reduce the sum over $s$ to Racah coefficients we write:
\begin{eqnarray}
(-1)^{\frac{1}{2}-s}=\langle s|\sigma_z|s \rangle=\sqrt{3} \, \mathcal{C}(\frac{1}{2} 1 \frac{1}{2};s,0,s) \, .
\end{eqnarray}
We make the permutation
\begin{eqnarray}
\mathcal{C}(\frac{1}{2} 1\frac{1}{2};s,M'-M,M'-M+s)=(-1) \,\mathcal{C}(1 \frac{1}{2} \frac{1}{2};M'-M,s,M'-M+s) \,,
\end{eqnarray}
and then write
\begin{eqnarray}
&&\mathcal{C}(1 \frac{1}{2} \frac{1}{2}; M'-M,s,M'-M+s) \,\mathcal{C}(\frac{1}{2} \frac{1}{2} J'; M'-M+s,M-s,M')\nonumber\\
&=&\sum_{j''}  \sqrt{2(2j''+1)} \,\mathcal{W}(1 \frac{1}{2} J' \frac{1}{2}; \frac{1}{2} j'') \,\mathcal{C}(\frac{1}{2} \frac{1}{2} j''; s,M-s,M)\, \mathcal{C}(1 j'' J'; M'-M, M, M') \,. ~~~~~~
\end{eqnarray}

By means of the former equation we have decoupled two CGC depending on $s$ into one depending on $s$ and one independent of $s$.  We can then combine the three $s$ dependent coefficients
in term of a Racah and find
\begin{eqnarray} \label{eq:s}
&&\sum_s \mathcal{C}(1 \frac{1}{2} \frac{1}{2}; 0,s,s) \,\mathcal{C}(\frac{1}{2} \frac{1}{2} J;s,M-s,M) \,\mathcal{C}(\frac{1}{2} \frac{1}{2} j'';s,M-s,M)\nonumber\\
&=&[2(2j''+1)]^{\frac{1}{2}}W(1 \frac{1}{2} J \frac{1}{2};\frac{1}{2} j'')\, \mathcal{C}(1 j'' J;0,M,M) \,.
\end{eqnarray}
Summing all the terms, we obtain
\begin{eqnarray}
N_{\mu}&=& AA'\left \{\big(1+B B'p^2(-1)^{M-M'}\big)(-1)^{1-J'} \sqrt{6} \, \sqrt{2J+1} \, \mathcal{W}(J \frac{1}{2} J' \frac{1}{2};\frac{1}{2} 1) \, \mathcal{C}(J 1 J';M,M'-M,M')  \right.  \nonumber\\
&-& 6\big(Bp+B'p(-1)^{M-M'}\big) \sum_{j''}(2j''+1)  \,\mathcal{W}(1 \frac{1}{2} J' \frac{1}{2};\frac{1}{2} j'')  \,\mathcal{W}(1 \frac{1}{2} J \frac{1}{2};\frac{1}{2} j'') \nonumber\\
&\times & \mathcal{C}(1 j''J';M'-M,M,M') \, \mathcal{C}(1 j''J;0,M,M) \bigg \}\delta_{\mu,M'-M}  \,.
\end{eqnarray}
The sum over $j''$ runs over $0,1$ as one can see from the CGC $\mathcal {C}(\frac{1}{2} \frac{1}{2} j'';s,M-s,M)$ of Eq.~\eqref{eq:s}

 We can apply this formula to the particular cases and we find

 \begin{itemize}
 \item[1)] $J=0,J'=0$
\begin{eqnarray} \label{eq:Nu00}
N_{\mu}= -A A'(B+B') p \,\delta_{\mu 0} \, \delta_{M 0} \, \delta_{M' 0} \, .
\end{eqnarray}
 \item[2)] $J=0,J'=1$
  \begin{eqnarray} \label{eq:Nu01}
N_{\mu} & =& A A'\left \{\big(1+B B'p^2(-1)^{-M'}\big)  \right. \nonumber\\
& +& \left. \sqrt{2} \, \big(B p+B'p(-1)^{-M'} \big) \, \mathcal{C}(1 1 1;M',0,M') \right\} \,\delta_{\mu M'} \,\delta_{M 0}  \, .
\end{eqnarray}
\item[3)] $J=1,J'=0$
  \begin{eqnarray} \label{eq:Nu10}
N_{\mu} & =& A A'\left\{ \big(1+ B B'p^2 (-1)^{M}\big) (-1)^{-M} \right. \nonumber\\
& -& \left. \sqrt{2} \, (-1)^{M}\, \big(Bp+B'p(-1)^{M}\big) \, \mathcal{C}(1 1 1;M,0,M) \right\} \,\delta_{\mu M} \,\delta_{M' 0}  \, .
\end{eqnarray}
The minus sign $-\sqrt{2}$  in  Eq.~\eqref{eq:Nu10} versus the $\sqrt{2}$ sign in Eq.~\eqref{eq:Nu01} looks surprising but turns to be irrelevant in $\overline{\sum}\sum|t|^2$
because,  as we shall show in next subsection, the two terms in Eqs.~\eqref{eq:Nu01} or \eqref{eq:Nu10} do not interfere when summing over polarizations.
\item[4)] $J=1,J'=1$
\begin{eqnarray} \label{eq:Nu11}
N_{\mu}& =&  AA'\left \{ \big(1+BB'p^2(-1)^{M-M'}\big) \,\sqrt{2}\, \mathcal{C}(1 11;M,M'-M,M')  \right. \nonumber\\
& -& \left. \big(B p+B'p(-1)^{M-M'}\big)  \, \delta_{M 0}\right. \nonumber\\
& -& \left. 2 \big(Bp+B'p(-1)^{M-M'}\big) \, \mathcal{C}(1 1 1;M'-M,M,M') \, \mathcal{C}(1 11;0,M,M)\right \}\delta_{\mu,M'-M} \,.~~~~~~~~~~
\end{eqnarray}
\end{itemize}

\section{Evaluation of $\overline{\sum} \sum |t|^2$}
As shown in  Eq. \eqref{eq:tt}   we must evaluate
\begin{equation}\label{eq:ttB}
\overline{\sum} \sum \left|t\right|^2= L^{00} M_0~ M^{*}_0 + L^{0i}   M_0 ~N^{*}_i + L^{i0}   N_i ~M_0^{*} + L^{ij}  N_i ~N_j^{*}  \, ,
\end{equation}
by taking the expressions for $M_0$, $N_{\mu}$ obtained above and $L^{\alpha\beta}$ from  Eq. \eqref{eq:ab}
\begin{eqnarray} \label{eq:Lab}
 {L}^{\alpha\beta} = 2~ \frac{p_\nu^\alpha p_l^\beta+p_l^\alpha p_\nu^\beta-p_\nu \cdot p_l g^{\alpha\beta}-i \epsilon^{\rho\alpha\sigma\beta} p_{\nu\rho} p_{l\sigma}}{m_\nu m_l} \, .
\end{eqnarray}
We make the calculation for each $J,J'$ combination, recalling that we are evaluating the matrix elements in the frame of the $\nu l$ at rest.\\
\begin{itemize}
\item[1)] $J=0,J'=0$
\begin{itemize}
\item[a)] First we evaluate the $L^{00}$ term of Eq.~\eqref{eq:ttB}.  The $\epsilon^{\rho\alpha\sigma\beta}$ term is zero since $\alpha=\beta=0$,
By using Eq. \eqref{eq:Lab} we find
\begin{eqnarray}\label{eq:t00}
L^{00} = \frac{m^2_l}{m_{\nu} m_l} \frac{M_{\rm inv}^{2(\nu l)}-m^2_l}{M_{\rm inv}^{2(\nu l)}} \, ,
\end{eqnarray}
and then summing over the polarizations $M,M'$ we find
\begin{eqnarray}
\overline{\sum_M}\sum_{M'} L^{00}M_0M_0^* = \frac{m^2_l}{m_{\nu} m_l} \frac{M_{\rm inv}^{2(\nu l)}-m^2_l}{M_{\rm inv}^{2(\nu l)}} (AA')^2 (1+B B' p^2)^2 \,.
 \end{eqnarray}
 \item[b)] The terms from $L^{0i}$,$L^{i0}$ are zero in the sum over polarizations and integration over phase space. Indeed, since  ${\bm{p}}_{\nu}=-{\bm{p}}_{l}$    in the $\nu l$ rest
frame.
\begin{eqnarray}
p_l^0 p_\nu^i + p_\nu^0 p_l^i =p_l^0 p_\nu^i - p_\nu^0 p_\nu^i    \,,
\end{eqnarray}
and
\begin{eqnarray}
\int d\Omega \,p_{\nu i} \rightarrow  \int d\Omega \,Y_{1 \mu}({\widehat{\bm{p}}}_{\nu})=0 \,.
\end{eqnarray}
The $\epsilon^{\rho\alpha\sigma\beta}$ term is now $\epsilon^{\rho 0 \sigma i}$ and
\begin{eqnarray}
\epsilon^{\rho 0\sigma i}p_{\nu \rho}p_{l \sigma}=-\epsilon^{\rho \sigma i} p_{\nu \rho}p_{l \sigma}=\epsilon^{\rho \sigma i} p_{\nu \rho} p_{\nu\sigma}=0 \,.
\end{eqnarray}
A similar approach can be used for the $\epsilon$ term in all the other cases and one can show that it always vanishes.
 \item[c)] For  the term from  $L^{ij} N_i N_j^*$, this evaluation is made easy recalling that we take ${\bm{p}}$  in the $z$ direction and we found in Eq.~\eqref{eq:Nu00} that $N_{\mu}$
is proportional to $\delta_{\mu0}$ $(\mu=0,i=3)$. Hence we get
\begin{eqnarray}
L^{33} N_3 N_3^*=L^{33} N_0 N_0^*   \, ,
\end{eqnarray}
and
\begin{eqnarray}
L^{33} &=& \frac{2}{m_{\nu}m_{l}} \left\{2 \widetilde{p}_{l 3} \widetilde{p}_{\nu 3} +\delta_{33} \widetilde{p}_l \cdot \widetilde{p}_{\nu}    \right\}  \\ \nonumber
 &=& \frac{2}{m_{\nu}m_{l}} \left\{ -2 \widetilde{p}_{\nu 3} \widetilde{p}_{\nu 3} + \widetilde{E}_{\nu} \widetilde{E}_{l}+ \widetilde{p}^2_{\nu}    \right\} \\ \nonumber
  &=& \frac{2}{m_{\nu}m_{l}} \left\{ - \frac{2}{3}\widetilde{p}^2_{\nu}   + \widetilde{E}_{\nu} \widetilde{E}_{l}+ \widetilde{p}^2_{\nu}    \right\}  \, ,
  \end{eqnarray}
where the last step uses the fact $\widetilde{p}_{\nu i}\widetilde{p}_{\nu j}$ becomes  $\frac{1}{3} \delta_{ij}\widetilde{p}^2_{\nu} $ upon integration over  $d\Omega ({\widehat{\bm{p}}}_{\nu})$
in the phase space.

Then we find
\begin{eqnarray}
\overline{\sum_M}\sum_{M'} L^{ij} N_i N_j^*=\frac{2}{m_{\nu}m_{l}}\big[\widetilde{E}_{\nu} \widetilde{E}_{l}+\frac{1}{3} \widetilde{p}_{\nu}^2\big](AA')^2 (B+B')^2 p^2,
\end{eqnarray}
and altogether
\begin{eqnarray}
\overline{\sum}\sum|t|^2&=&\frac{(AA')^2}{m_{\nu}m_l} \left\{\frac{m^2_l (M_{\rm inv}^{2(\nu l)}-m^2_l)}{M_{\rm inv}^{2(\nu l)}} (1+B B' p^2)^2 \right. \\ \nonumber
 &+& \left. 2(\widetilde{E}_{\nu}\widetilde{E}_l+\frac{1}{3}\widetilde{p}_{\nu}^2)(B+B')^2p^2 \right\} \,.
\end{eqnarray}
\end{itemize}
\item[2)] $J=0,J'=1$
\begin{itemize}
\item[a)] Using   Eqs.~\eqref{eq:M01} and \eqref{eq:t00}, we find
\begin{eqnarray}
\overline{\sum}\sum L^{00}M_0M_0^* = \frac{m^2_l}{m_{\nu} m_l} \frac{M_{\rm inv}^{2(\nu l)}-m^2_l}{M_{\rm inv}^{2(\nu l)}} (AA')^2 (B +B')^2 p^2 \,.
 \end{eqnarray}
 \item[b)] The $\epsilon$  term and $L^{0i}$ also vanish when integrating over $d\Omega ({\widehat{\bm{p}}}_\nu)$.
 \item[c)] For the term from  $L^{ij} N_i N_j^*$ we need  Eq.~\eqref{eq:Nu01} and
 \begin{eqnarray} \label{eq:Lij}
L^{ij}=\frac{2}{m_{\nu} m_l} \left\{ -2 p_{\nu}^{i} p_{\nu}^{j} + \delta_{i j}(p_{\nu}\cdot p_l) \right\}  \,.
\end{eqnarray}
We now write in spherical basis
\begin{eqnarray}
\sum_ip_{\nu i}N_i&=&\sum_{\alpha}(-1)^{\alpha}p_{\nu \alpha}N_{-\alpha} \, ,\\\nonumber
\sum_ip_{\nu i}N_i^*&=&\sum_{\beta}p_{\nu \beta}N_{\beta}^* \, .
\end{eqnarray}
One evaluates first the $\widetilde{p}_{\nu i} \widetilde{p}_{\nu j}$ term, writes
\begin{eqnarray}
\widetilde{p}_{\nu \alpha}=\sqrt{\frac{4\pi}{3}}\tilde{p}_{\nu} Y_{1 \alpha }({\widehat{\bm{p}}}_\nu)  \, ,
\end{eqnarray}
and use that
\begin{eqnarray}
\int Y_{1 \alpha }({\widehat{\bm{p}}}_\nu)  Y_{1 \beta}({\widehat{\bm{p}}}_\nu) d\Omega ({\widehat{\bm{p}}}_\nu) = \delta_{\alpha \beta}   \, ,
\end{eqnarray}
and we get
\begin{eqnarray}
&&\overline{\sum_M} \sum_{M'} -\frac{4}{m_{\nu}m_l}\tilde{p}^i_{\nu}\tilde{p}^j_{\nu} N_i N_j^* =-\frac{4}{m_{\nu}m_l} (A A')^2 \frac{1}{3}\widetilde{p}^2_{\nu}  \,\delta_{M 0} \\ \nonumber
 &\times&
\left\{\big(1+B B'p^2(-1)^{-M'}\big)+\sqrt{2} \, \big(B p+B'p(-1)^{-M'} \big) \, \mathcal{C}(1 1 1;M',0,M')\right\}^2   \, .
\end{eqnarray}
And explicit evaluation of the square of the bracket and the sum over $M,M'$ gives at the end
\begin{eqnarray}
&&\overline{\sum_M} \sum_{M'} -\frac{4}{m_{\nu}m_l}\tilde{p}^i_{\nu}\tilde{p}^j_{\nu} N_i N_j^* \\ \nonumber
&=&-\frac{4}{m_{\nu}m_l} (A A')^2 \frac{1}{3}\widetilde{p}^2_{\nu}  \left\{3-6 B B'p^2 +2(B^2+B'^2) p^2 +3 (B B' p^2)^2 \right\}  \, .
\end{eqnarray}
The term $\delta_{ij}(p_{\nu}\cdot p_{\ell}) N_i N_j^*$ is evaluated in the same way and we find
\begin{eqnarray}
&&\overline{\sum_M} \sum_{M'} L^{ij} N_i N_j^* \\ \nonumber
&=&\frac{2}{m_{\nu}m_l} (A A')^2 \big(\widetilde{E}_{\nu}\widetilde{E}_l+\frac{1}{3}\tilde{p}_{\nu}^2 \big) \left\{3-6 B B'p^2 +2(B^2+B'^2) p^2 +3 (B B' p^2)^2 \right\}  \, ,
\end{eqnarray}
and summing the $L^{00} M_0 M_0^*$ contribution  we obtain at the end
\begin{eqnarray}
\overline{\sum_M}\sum_{M'}|t|^2 &=&\frac{(AA')^2}{m_{\nu} m_l} \left\{2\big(\widetilde{E}_{\nu}\widetilde{E}_l+\frac{1}{3}\tilde{p}_{\nu}^2 \big) \left[3-6 B B'p^2 + \right. \right.  \\ \nonumber
& & \left. \left. 2(B^2+B'^2) p^2  +3 (B B' p^2)^2 \right] +\frac{m^2_l (M_{\rm inv}^{2(\nu l)}-m^2_l)}{M_{\rm inv}^{2(\nu l)}} (B+B')^2 p^2 \right\}  \, .~~~~~~
\end{eqnarray}
\end{itemize}
 \item[3)] $J=1,J'=0$ \\
 We obtain the same result as before, but must multiply by $\frac{1}{3}$ to take into account the average over the initial polarizations.\\
\item[4)] $J=1,J'=1$
\begin{itemize}
\item[a)] $L^{00} M_0 M^*_0$  term \\
We  need the expression of Eq.~\eqref{eq:M11}.  We can see that in $M_0 M_0^*$ we have terms like
\begin{eqnarray}
\sum_M {\cal C}(1 1 1;M,0,M)&=&0   \,, \nonumber\\
\sum_M {\cal C}(1 1 1;M,0,M) \, {\cal C}(1,1,1;M,0,M)&=& \frac{1}{2}+0+\frac{1}{2}=1  \,,
\end{eqnarray}
then
\begin{eqnarray}
\overline{\sum_M} \sum_{M'} L^{00} M_0 M_0^*= (AA')^2 \frac{m^2_l}{m_{\nu} m_l} \frac{M_{\rm inv}^{2(\nu l)}-m^2_l}{M_{\rm inv}^{2(\nu l)}}
\left\{(1+BB'p^2)^2 +\frac{2}{3}(B+B')^2 p^2 \right\}   \,.  ~~~~~~~~~~
\end{eqnarray}
\item[b)]For the same reasons as before $L^{0i}M_0N_i^*$,$L^{i0}N_iM_0^*$ vanish in the integration over phase space.
\item[c)] $L^{ij} N_i N_j^*$  term \\
$L^{ij}$ is given by Eq.~\eqref{eq:Lij}, proceeding as we have done before.  The extra CGC in this term are easily handled since
$\mathcal{C}(1 1 1;0,0,0)$ and the other coefficients are all $\frac{1}{\sqrt{2}}$  except for a phase.  We need the expression of Eq.~\eqref{eq:Nu11} and we
can easily see that in the sum over $M,M'$ the three terms of this equation do not interfere.
An explicit bookkeeping  and some algebra allows as to write finally
\begin{eqnarray}
\overline{\sum_M}\sum_{M'}|t|^2 &=& \frac{1}{3} \,\frac{(A A')^2}{m_{\nu} m_l} \left\{ \frac{ 3\, m^2_l(M_{\rm inv}^{2(\nu l)}-m^2_l)}{M_{\rm inv}^{2(\nu l)}}  \left[(1+BB'p^2)^2+ \frac{2}{3}(B+B^{\prime})^2 p^2 \right] \right. \\ \nonumber \,
&+& \left. 2\left(\widetilde{E}_\nu \widetilde{E}_l + \frac{1}{3}\widetilde{p}^2_\nu \right) \left[6 + 7(B^2+ B'^2)p^2 - 4 B B' p^2+6 (BB'p^2)^2 \right] \right\}  \,. ~~~~~~~~
 \end{eqnarray}
\end{itemize}
\end{itemize}


\begin{thebibliography}{99}

\bibitem{browder}
  T.~E.~Browder and K.~Honscheid,
  Prog.\ Part.\ Nucl.\ Phys.\  {\bf 35}, 81 (1995).


\bibitem{isgur}
  N.~Isgur and M.~B.~Wise,
  Phys.\ Lett.\ B {\bf 232}, 113 (1989).

\bibitem{isgur2}
  N.~Isgur, D.~Scora, B.~Grinstein and M.~B.~Wise,
  Phys.\ Rev.\ D {\bf 39}, 799 (1989).

\bibitem{Wirbel:1988ft}
  M.~Wirbel,
  Prog.\ Part.\ Nucl.\ Phys.\  {\bf 21}, 33 (1988).

\bibitem{neubert}
  M.~Neubert and B.~Stech,
  Adv.\ Ser.\ Direct.\ High Energy Phys.\  {\bf 15}, 294 (1998).

\bibitem{ecker}
  G.~Ecker,
  Prog.\ Part.\ Nucl.\ Phys.\  {\bf 35}, 1 (1995).


\bibitem{neubert2}
  M.~Neubert,
  Int.\ J.\ Mod.\ Phys.\ A {\bf 11}, 4173 (1996).

\bibitem{Antonelli:2009ws}
  M.~Antonelli {\it et al.},
  Phys.\ Rept.\  {\bf 494}, 197 (2010).

\bibitem{Fajfer} S. Fajfer, J.F. Kamenik, and I. Ni$\check{\rm s}$and$\check{\rm z}$i$\acute{\rm c}$, Phys. Rev. D {\bf 85}, 094025 (2012).

\bibitem{German}
Xiao-Gang He, German Valencia,  Phys.  Lett.  B  {\bf 779}, 52 (2018).


\bibitem{nieves}
C. Albertus, E. Hern$\acute{\rm a}$ndez,  J. Nieves, and J. M. Verde-Velasco, Phys.  Rev.  D.  {\bf 71}, 113006 (2005).

\bibitem{wangwang}
 Tianhong Wang, Yue Jiang, Tian Zhou, Xiao-Ze Tan, and Guo-Li Wang,   arXiv:1804.06545 [hep-ph].

\bibitem{changyang}
Q. Chang, J. Zhu, X. L. Wang, J. F. Sun, and Y. L. Yang, Nucl. Phys. B  {\bf 909}, 921 (2016).


\bibitem{Neubert}
Matthias Neubert,  Phys.  Rept.   {\bf 245}, 259  (1994).

\bibitem{Manohar}
  A.~V.~Manohar and M.~B.~Wise,
  Camb.\ Monogr.\ Part.\ Phys.\ Nucl.\ Phys.\ Cosmol.\  {\bf 10}, 1 (2000).



\bibitem{Navarra}
Fernando S. Navarra, Marina Nielsen, Eulogio Oset, and Takayasu Sekihara,  Phys.  Rev.  D.  {\bf 92}, 014031 (2015).

\bibitem{mandl}
F. Mandl and G. Shaw, Quantum Field Theory, John Wiley $\&$ Sons, (1984).


\bibitem{Itzykson}
C.~Itzykson and J.~B.~Zuber,  Quantum Field Theory, Mecraw-Hill, 1980.

\bibitem{pdg}
 C. Patrignani et al. (Particle Data Group). Chin. Phys. C {\bf 40}, 100001 (2016).

\bibitem{HFLAV} Y. Amhis  {\sl et al.} Heavy Flavor Averaging Group (HFLAV), Eur. Phys. J. C {\bf 77}, 895 (2017).

\bibitem{lhcbd}
R. Aaij {\sl et al.} (LHCb Collaboration), Phys. Rev. D {\bf 97}, 072013 (2018).

\bibitem{lat1}
Jon A. Bailey {\sl et al.} (Fermilab Lattice and MILC Collaborations),  Phys. Rev. D {\bf 92}, 034506 (2015).

\bibitem{lat2}
Heechang Na {\sl et al.} (HPQCD Collaboration), Phys. Rev. D {\bf 92}, 054510 (2015); Erratum Phys. Rev. D {\bf 93}, 119906 (2016).

\bibitem{Genaro}
J. E. Chavez-Saab, Genaro Toledo, arXiv:1806.06997

\bibitem{lhcbj}
R. Aaij {\sl et al.} (LHCb Collaboration),  Phys. Rev. Lett.  {\bf 120}, 121801  (2018).

\bibitem{Semay}
A. Yu. Anisimov, I. M. Narodetskii, C. Semay, B. Silvestre-Brac,  Phys.  Lett.  B {\bf 452}, 129  (1999).

\bibitem{Kiselev} 
V. V. Kiselev, arXiv:hep-ph/0211021

\bibitem{Ivanov}
Mikhail A. Ivanov, J\"{u}rgen G. K\"{o}rner, and Pietro Santorelli,  Phys. Rev. D {\bf 73}, 054024 (2006).

\bibitem{eli}
E. Hern$\acute{\rm a}$ndez, J. Nieves, and J. M. Verde-Velasco,  Phys. Rev. D {\bf 74}, 074008 (2006).

\bibitem{zhangwang} J. M. Zhang and G. L. Wang, Chin. Phys. Lett. {\bf 27}, 051301 (2010).

\bibitem{rose}
M. E. Rose, Elementary Theory of Angular Momentum, John Wiley $\&$ Sons, 1957.



\bibitem{ebert11} D. Ebert, R. N. Faustov, and V. O. Galkin, Eur. Phys. J. C {\bf 71}, 1825 (2011).

\bibitem{liangoset}
  W.~H.~Liang and E.~Oset,  arXiv:1804.00938 [hep-ph] [Eur. Phys. J. C in print).
\bibitem{daioset}
  L.~R.~Dai, R.~Pavao, S.~Sakai and E.~Oset,
  arXiv:1805.04573 [hep-ph].

\end{thebibliography}
\end{document}